\documentclass[preprint2]{aastex62}

\graphicspath{{./}{figures/}}

\received{}
\revised{}
\accepted{}
\submitjournal{ApJ}

\shorttitle{Coronal loop transverse oscillations}
\shortauthors{Afanasyev et al.}

\begin{document}

\title{Coronal loop transverse oscillations excited by different driver frequencies}

\correspondingauthor{Andrey Afanasyev}
\email{andrei.afanasev@kuleuven.be}

\author{Andrey Afanasyev}
\affiliation{Centre for mathematical Plasma Astrophysics, KU Leuven \\
Celestijnenlaan 200B, box 2400, B-3001 Leuven, Belgium}
\affiliation{Institute of Solar-Terrestrial Physics of SB RAS, Irkutsk, Russia}

\author{Konstantinos Karampelas}
\affiliation{Centre for mathematical Plasma Astrophysics, KU Leuven \\
Celestijnenlaan 200B, box 2400, B-3001 Leuven, Belgium}

\author{Tom Van Doorsselaere}
\affiliation{Centre for mathematical Plasma Astrophysics, KU Leuven \\
Celestijnenlaan 200B, box 2400, B-3001 Leuven, Belgium}



\begin{abstract}

We analyse transverse oscillations of a coronal loop excited by continuous monoperiodic motions of the loop footpoint at different frequencies in the presence of gravity. 
Using the MPI-AMRVAC code, we perform three-dimensional numerical magnetohydrodynamic simulations, considering the loop as a magnetic flux tube filled in with denser, hotter, and gravitationally stratified plasma. We show the resonant response of the loop to its external excitation and analyse the development of the Kelvin-Helmholtz  instability at different heights. We also study the spatial distribution of plasma heating due to transverse oscillations along the loop. The positions of the maximum heating are in total agreement with those for the intensity of the Kelvin-Helmholtz instability, and correspond to the standing wave anti-nodes in the resonant cases. The initial temperature configuration and plasma mixing effect appear to play a significant role in plasma heating by transverse footpoint motions. In particular, the development of the Kelvin-Helmholtz instability in a hotter loop results in the enhancement of the mean plasma temperature in the domain. 

\end{abstract}

\keywords{magnetohydrodynamics (MHD) --- waves --- 
Sun: corona --- Sun: oscillations}


\section{Introduction} \label{sec:intro}

Observations of the Sun in Extreme-Ultraviolet (EUV) spectral lines show that wave and oscillatory plasma motions exist everywhere throughout the solar corona (e.g., \citealp{Tomczyk2007}, see also reviews by \citealp{DeMoortel_Nakariakov, Liu2014}). Traditionally, waves in the solar corona are considered as one of the possible candidates to heat the plasma (e.g., \citealp{Arregui20140261, DeMoortel_Browning9}). Observational studies of coronal wave dynamics provide estimates for the total energy flux able to reach the corona from the chromosphere to be about $170$~W~m$^{-2}$ for the incompressible transverse motions and $460$~W~m$^{-2}$ for compressive motions (e.g., \citealp{Morton2012}, see also \citealp{DePontieu2007, Tomczyk2009ApJ...697.1384T, Thurgood2014ApJ...790L...2T}), which is comparable to the coronal heating requirements of about $300$~W~m$^{-2}$ \citep{Withbroe1977}. Despite the disappointing conclusions of observation-based simulations reporting that wave heating calculated within the scope of the known dissipative mechanisms does not appear to be sufficient to balance the radiative losses (e.g., \citealp{pagano2018,pagano2017, PaganoDeMoortel2019}), it is generally accepted that wave heating at least contributes some part into the general energetic budget of the solar corona. Moreover, one should take into account that both observations and simulations are still quite far from their desired state. Coronal instruments do not allow observers to resolve distances of about ten kilometre size to check results of up-to-date numerical magnetohydrodynamic (MHD) simulations. In turn, MHD simulations cannot be performed with the resolution that allows setting real (small) values of viscosity and resistivity.

Dynamics of coronal waves can be described satisfactorily within certain limits in the single-fluid MHD approximation. In particular, transverse oscillations of coronal loops were interpreted in terms of MHD kink modes of cylindrical plasma waveguides \citep{Aschwanden1999ApJ...520..880A,Nakariakov1999Sci...285..862N}. Analysis of SDO/AIA data showed statistical properties of the observed transverse oscillations of loops \citep{Zimovets2015,Goddard2016A&A...585A.137G}. Along with these initially detected transverse oscillations, which have relatively high amplitude and decay fast (in several oscillation cycles), recently the so-called decay-less regime of small-amplitude transverse oscillations was discovered \citep{Wang2012-L27,Nistico2013,Anfinogentov2015}. Decay-less oscillations were interpreted in terms of standing kink waves driven continuously at the loop footpoints \citep{Karampelas2019,Guo2019}. In the other interpretation, they were considered as a self-oscillatory process due to the interaction of the loops with quasi-steady flows \citep{Nakariakov2016}. Recently, \citet{Duckenfield2018ApJ...854L...5D} reported on the first detection of multiple harmonics in decay-less kink oscillations, while for decaying transverse oscillations, harmonics were detected by several authors (e.g., \citealp{VD2007,DeMoortelBrady2007,HongboLi0004-637X-842-2-99}). The multi-harmonic structure of observed loop oscillations implies that the driver of oscillations is more complicated, which should be taken into account in simulations. In addition, these observational findings provide promising tools for seismological techniques to determine plasma parameters (e.g., \citealp{Stepanov_book_2012}). 

The prevailing mechanism responsible for the rapid damping of loop transverse oscillations is resonant absorption of kink waves (e.g., \citealp{ruderman2002,goossens2002,pascoe2010,pascoe2012,hood2013}), although other interpretations were also proposed, for instance, dispersive evolution of an external pulse exciting loops \citep{uralov2003AstL...29..486U,Terradas2005-dispersion}. In the resonant absorption mechanism, azimuthal Alfv\'en waves are generated at the non-uniform loop boundary due to the mode coupling of kink waves (which are in fact of fast-magnetoacoustic nature) and Alfv\'en waves. This leads to energy transfer to the Alfv\'en mode, and attenuation of kink oscillations. In their turn, generated Alfv\'en waves experience phase mixing due to non-uniformity of the loop boundary, creating small-scale shear flows and heating the plasma (e.g., \citealp{heyvaerts1983A&A...117..220H, pascoe2010,pagano2017,pagano2018,PaganoDeMoortel2019}).

Shear plasma flow may lead to the development of the Kelvin-Helmholtz instability (KHI). The unstable state of phase-mixed Alfv\'en waves in magnetic structures with non-uniform layers was theoretically studied by \citet{heyvaerts1983A&A...117..220H} and \citet{browning1984A&A...131..283B}. \citet{ofman1994GeoRL..21.2259O} and \citet{ofman1995JGR...10023427O} modelled resonant excitation of Alfv\'en waves in the slab geometry, paying special attention to nonlinear effects and KHI, which led to the turbulent enhancement of plasma dissipative parameters. Recently, \citet{Barbulescu2019} and \citet{Hillier2019} analytically analysed conditions for the KHI development at the boundary layer of a magnetic tube.

On the other hand, numerical simulations showed that KHI can be induced by transverse oscillations of loops (e.g., \citealp{Terradas2008,Antolin2014,antolin2016,Terradas2018}). \citet{Terradas2008} demonstrated that KHI exists both in the case of a sharp and smooth boundary layer. The strong turbulent effect due to the instability affects the loop structure, deforming it significantly \citep{Magyar2016_multi-stranded,Karampelas2018}. \citet{Antolin2014,antolin2016,antolin2017} analysed the effects of KHI in observations of coronal loops, comparing synthetic EUV images to observations. In particular, in \citet{antolin2016}, the connection between decay-less oscillations and line-of-sight effects due to KHI vortices was studied, as well as their seismological applications. \citet{Magyar2016damping_of_standing} considered large-amplitude standing kink oscillations, showing the nonlinear contribution of KHI into their damping rate. Recently, \citet{Karampelas2017} analysed the temperature evolution of oscillating coronal loops and obtained an increase in the volume averaged temperature in the loop vicinity. They found manifestations of plasma heating both due to viscosity and resistivity, however, concluded that the heating effect due to the KHI mixing of plasmas of different temperatures was dominant and could hide effects of heating due to dissipation of energy of small-scale plasma flows.

Previous studies considered mainly the continuous monoperiodic driver excitation of a coronal loop at the fundamental standing wave frequency, although \citet{pagano2017} and \citet{pagano2018} analysed standing wave regimes including several harmonics. However, it seems to be unlikely that real drivers correspond exactly to the fundamental oscillatory modes of loops. Moreover, in the case of gravitational stratification of the plasma, the exact values of the loop eigenmodes are unknown and can only be estimated in some approximation \citep{Andries2005, Magyar2015}. Very recently, \citet{PaganoDeMoortel2019} studied coronal loop oscillations triggered by random footpoint motions with the observed coronal power spectrum, focusing on the heating effect of phase mixing mechanism. The aim of this study is to analyse the response of a stratified coronal loop to its excitation by a driver at different frequencies, find the resonant frequencies of the loop, and analyse the height dependence of plasma heating by transverse waves. 

The paper is organised as follows. In Sect.~\ref{sec:model} details of the model and method used as well as the simulation setup are described. We present results of our simulations and discuss them in Sect.~\ref{sec:results}. In Sect.~\ref{sec:summary}, we summarise our findings.

\section{Model and Method} \label{sec:model}

Using the MPI-AMRVAC code \citep{xia2018}, we perform three-dimensional (3D) numerical simulations of the propagation of transverse MHD waves in a gravitationally stratified coronal loop. We consider a density and temperature enhanced magnetic flux tube in a uniform straight magnetic field directed along the $z$-axis in our setup. In order to take into account the curvature of coronal loops, the gravitational acceleration is modulated as (under the assumption of a semi-circular loop shape),
\begin{equation}
\label{gravity_law}
g\left( z \right) = g_{\odot} \cos \left( \pi z / L \right),
\end{equation}
where $g_{\odot}$ is the surface value of the solar gravity acceleration, $L=200$~Mm is the loop length, so we have the maximum gravity value at the loop footpoints, and no gravity at the loop apex. Note that, using this model, we can only analyse loop oscillations of the horizontal polarisation in the low-$\beta$ approximation, otherwise, Eq.~\ref{gravity_law} is not applicable.

The plasma temperature in the volume is set as
\begin{eqnarray}
    T = T_e + \left( T_i - T_e \right) \zeta \left( x, y \right), \label{temperature_law} \\
\zeta \left( x, y \right) = \frac{1}{2} \left( 1 - \tanh  \left[ \left( \frac{r}{R} - 1 \right) b \right] \right), \label{geometrical_factor}
\end{eqnarray}
where indices $i$ and $e$ denote values inside and outside the loop, respectively, $\zeta$ is the factor describing the loop and its non-uniform boundary, $b=16$ is the ratio of the loop radius, $R=1$~Mm, to the characteristic size of the loop boundary, $r=\sqrt{x^2+y^2}$ is the distance from the loop centre in the $xy$-planes.

The equation of hydrostatic equilibrium in the $z$-direction combined with Eq.~\ref{gravity_law}, Eq.~\ref{temperature_law}, and Eq.~\ref{geometrical_factor} give the plasma density profile, provided that the density variation $\rho_0 \left( x, y \right)$ in the $xy$-planes has been defined,
\begin{eqnarray}
\rho_0 = \rho_e + \left( \rho_i - \rho_e \right) \zeta \left( x, y \right), \label{2d_density_law} \\
\rho = \rho_0 \exp \left( - \frac{L g_{\odot} \sin \left( \pi z/L \right) }{\pi \alpha T} \right), \label{density_law}
\end{eqnarray}
where $\alpha$ is a constant determined from the equation of gas state. 

\begin{figure*}
\gridline{\fig{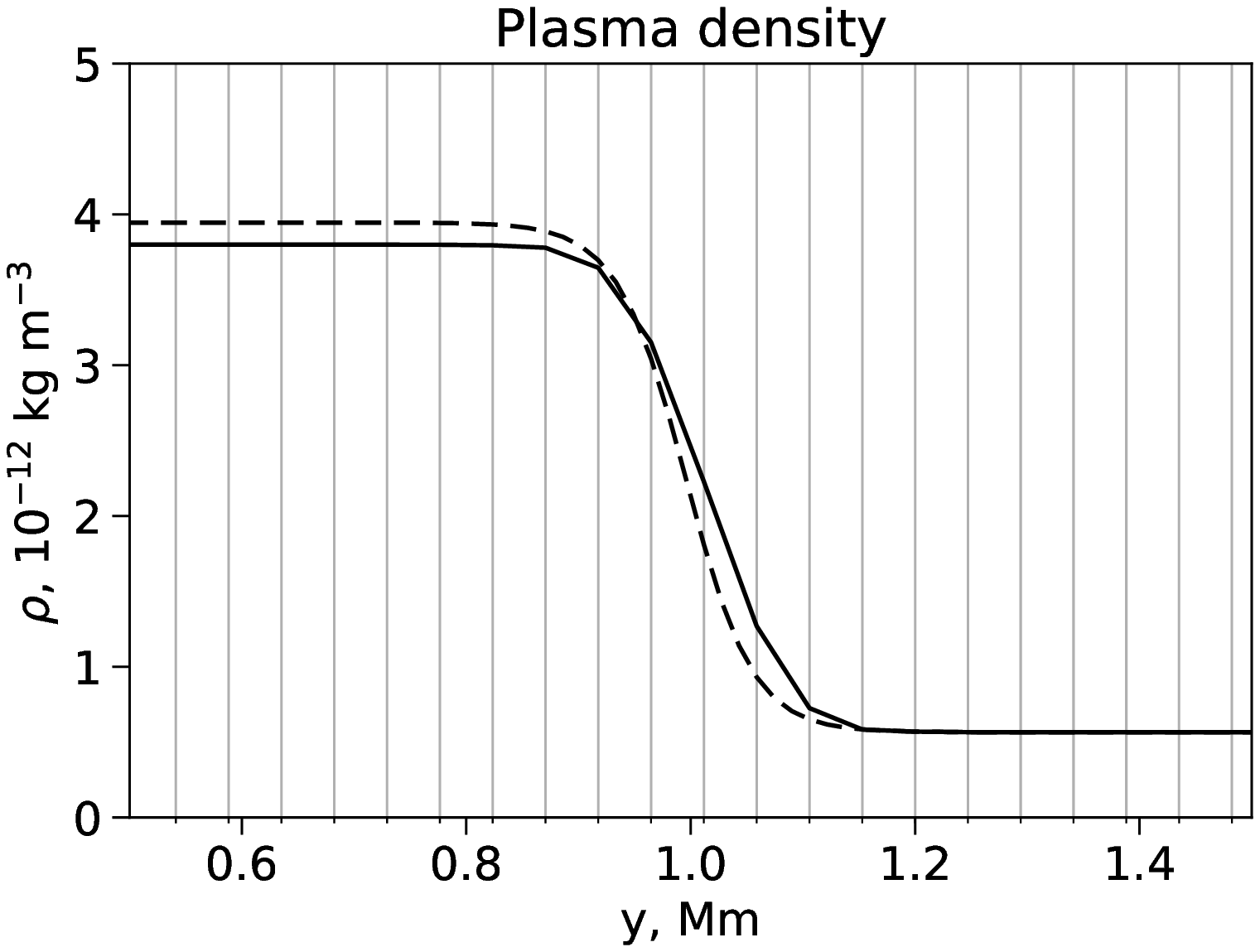}{0.4\textwidth}{(a)}
          \fig{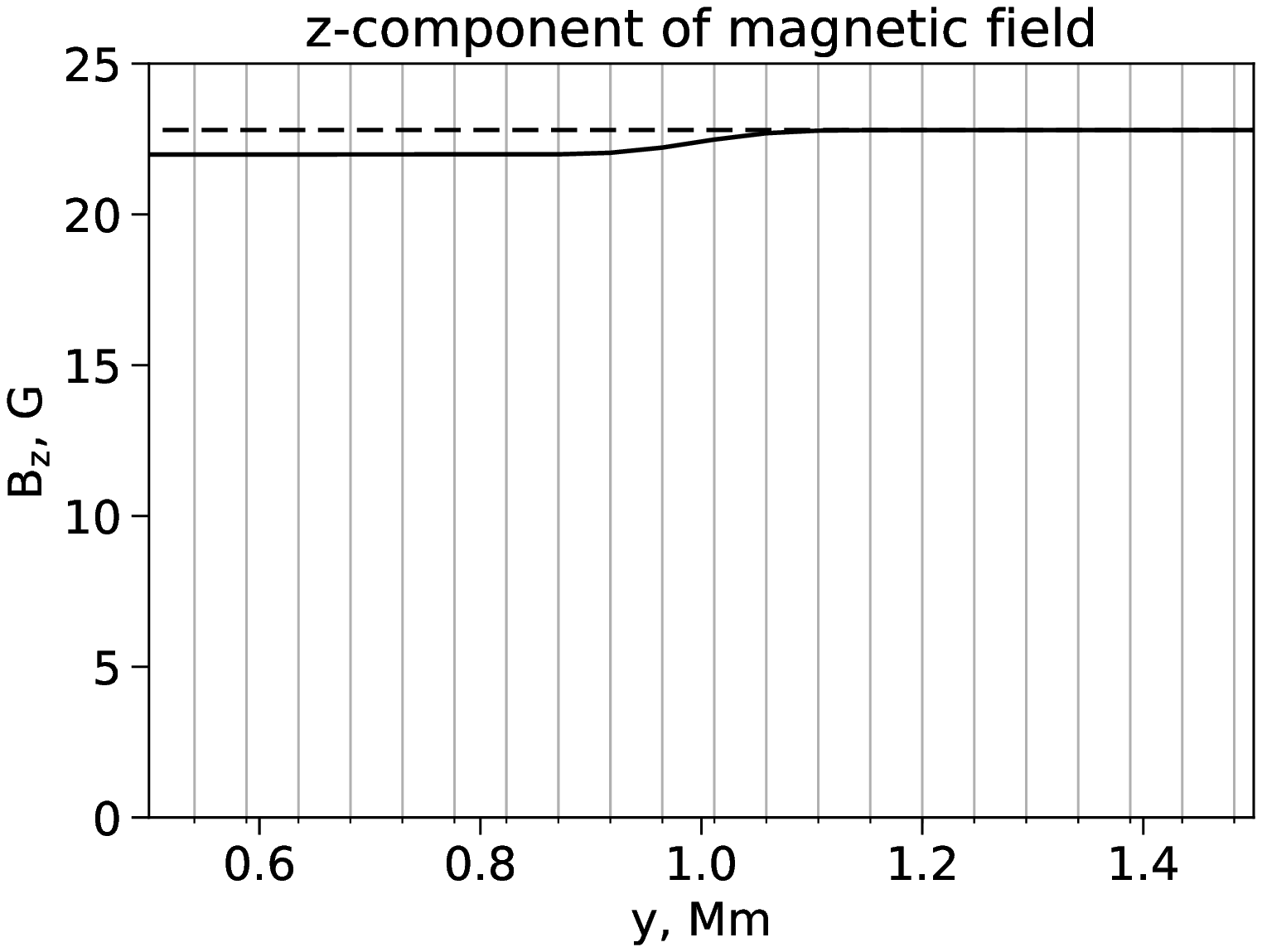}{0.4\textwidth}{(b)}
          }
\caption{Density (a) and $z$-component of the magnetic field (b) near the loop boundary at the loop apex. The dashed lines show the initial unstable state defined analytically, the solid lines show the state after the rapid relaxation process. The vertical lines represent the size of cells in simulations. \label{fig_ini_state}}
\end{figure*}

This configuration is obviously out of MHD equilibrium in the $xy$-planes. Moreover, due to gravitational stratification of the plasma, it is difficult to balance gas and magnetic pressure in these planes at the loop boundary at all heights. However, since we are interested in the low-$\beta$ case, it is quite reasonable first to run a simulation to let the initial configuration reach the equilibrium state. The relaxation of the initial state occurs very rapidly, compared to the periods of transverse waves (see also \citealp{Magyar2015}). However, it changes slightly the initial configuration, in particular reducing the magnetic field strength inside the loop. Figure~\ref{fig_ini_state} shows the difference in the initial and balanced plasma density and magnetic field at the loop apex near the loop boundary. Also, the new state includes propagating waves of slow magnetoacoustic nature of several kilometres per second amplitude, which are trapped between the top and bottom boundaries of the computational domain. We keep this in mind when analysing results of simulations. 
The difference in the total mass of the plasma inside the whole computational domain for the initial and equilibrium states is about $0.5\%$, while the temporal analysis of the total plasma mass shows that it is conserved within $0.15\%$ during the simulation runtime. 

The waves are driven by transverse displacements of the tube footpoint at $z=0$. We take the idea of a continuous monoperiodic dipole-like driver (\citealp{Karampelas2017,Karampelas2018}, see also \citealp{pascoe2010}), setting the uniform value of the $y$-component of the plasma velocity $v_y$ inside the loop, and dipole-like flows outside. The velocity amplitude of the driver is $5$~km~s$^{-1}$ for all simulations. We consider different driver frequencies to study the behaviour of the loop in response to its excitation at one of its footpoints.

Due to the obvious setup symmetry with respect to the $x=0$~plane, we model only half a loop that allows us to reduce the use of computational resources. At this plane, we set symmetric boundary conditions for all variables, except for the $x$-components of the momentum and magnetic field, which have anti-symmetric conditions. The other lateral boundaries are open to let perturbations leave the computational domain. At the top boundary, we set anti-symmetric conditions for the momentum to get the reflection of waves and tie the field lines, and prevent plasma flows through the boundary. At the bottom boundary, the driver regime discussed is implemented. We use open boundary conditions for the magnetic field components and extrapolate the barometric equilibrium for pressure and density into the ghost cells at the top and bottom boundaries.

We solve numerically 3D ideal MHD equations with MPI-AMRVAC code with the one-step TVD method with the Woodward limiter, which is implemented in the code. We use a uniform grid $128 \times 256 \times 64$ cells for most of the simulations. The domain size is $((0,6), (-6,6), (0,200))$~Mm, so we have a resolution of about 47~km per pixel in the $xy$-planes, and 3.17~Mm per pixel in the $z$-direction, assuming that the loop evolution in that direction is much smoother, which is confirmed by results of previous studies (e.g., \citealp{Terradas2008,Antolin2014,Karampelas2017}). For convergence tests, we perform simulations  on a grid of $88 \times 352 \times 64$ cells in a domain of $((0,2), (-4,4), (0,200))$~Mm size for two frequencies of the driver, which allows us to obtain a double resolution of about 23~km per pixel in the $xy$-planes.

The following initial parameters are chosen for the simulations: $T_e = 1$~MK, $T_i = 3$~MK, the magnetic field strength is 22.8~G, the plasma number density at the top and bottom boundaries inside the loop is $3.0 \times 10^9$~cm$^{-3}$, and $1.0 \times 10^9$~cm$^{-3}$ outside it. The sound speed inside the loop is about 260~km~s$^{-1}$, while the Alfv\'en speed inside the loop is about 1005~km~s$^{-1}$ at the apex, and 810~km~s$^{-1}$ at the footpoints. Respectively, the characteristic values of plasma beta are about $0.06$ and $0.10$.  We run simulations for 2000~s, which exceeds six fundamental standing mode periods ($\approx 328$~s) as determined from the simulation results. Note that estimates of the fundamental standing mode period for the stratified loop with the use of a weight function (see \citealp{Andries2005}) give very close value of 310~s, and a weighted mean kink speed of 1290~km~s$^{-1}$. For the second and third harmonics, one can obatin values of 162~s and 108~s, using the same approach.  Periods of the footpoint driver in our simulations are in a range of 92--421~s.

\section{Simulation results and Discussion} \label{sec:results}

We perform a set of MHD simulations of driven coronal loop oscillations, setting different values of the footpoint driver period. The loop behaves as a resonator with the spectrum of eigenfrequencies. From the general theory of oscillations, it is known that if the frequency of the driver corresponds to one of the eigenfrequencies, the system comes into resonance, so we can detect the growth of oscillations. We characterise the magnitude of oscillations by the displacement of the centre of mass of the loop in each horizontal cross-section as well as by its speed. The position of the loop centre of mass depends on the domain, over which we calculate its value. 
In particular, taking into account the turbulent KHI eddies near the loop boundary as well as a contribution from the neighbourhood plasma can significantly decrease the calculated value of the displacement of the loop. In order to diminish those contributions, we choose different threshold values of the plasma density, $\rho_{\mathrm{th}}=$~0.5, 0.7, and 0.9, with respect to some reference value at a given height. The reference value of the plasma density is calculated as a mean value inside the narrow cylinder, $r<0.5$~Mm, at a given height from the initial state (clearly, from that after relaxation, see Sect.~\ref{sec:model}), which corresponds to the simulation snapshot just before starting the footpoint driver.

For resonant frequencies, after the time required for kink waves to reach the opposite footpoint and return back, a standing-wave regime is established in the loop. In fact, this regime is not a proper standing wave, even for resonant frequencies, because the energy of oscillations is continuously delivered from the driver, so occasionally oscillations also appear at the nodes of the standing wave. The location of the nodes and anti-nodes depends on the driver frequency, as expected from the general theory.

\begin{figure*}
\gridline{\fig{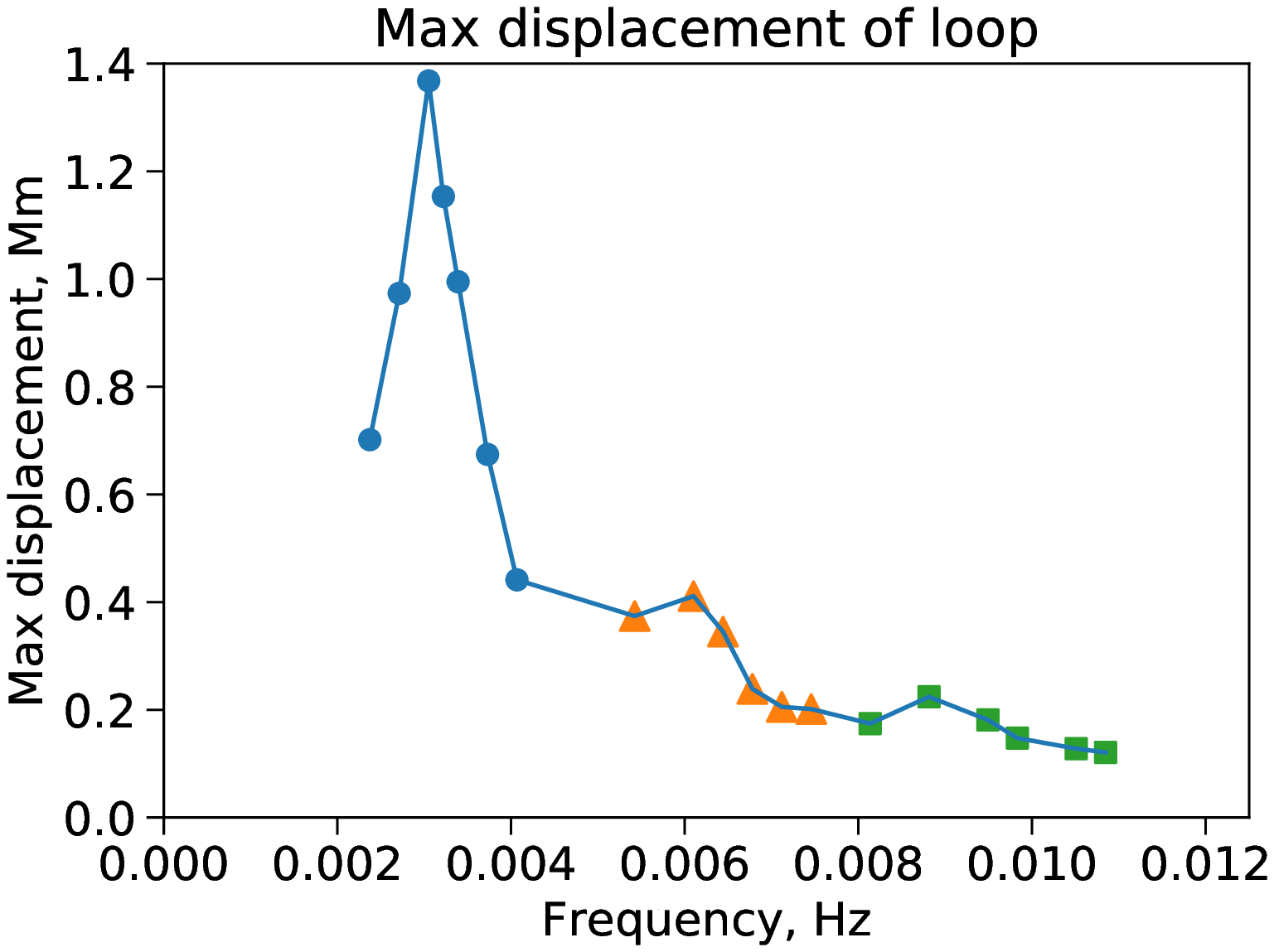}{0.4\textwidth}{(a)}
          \fig{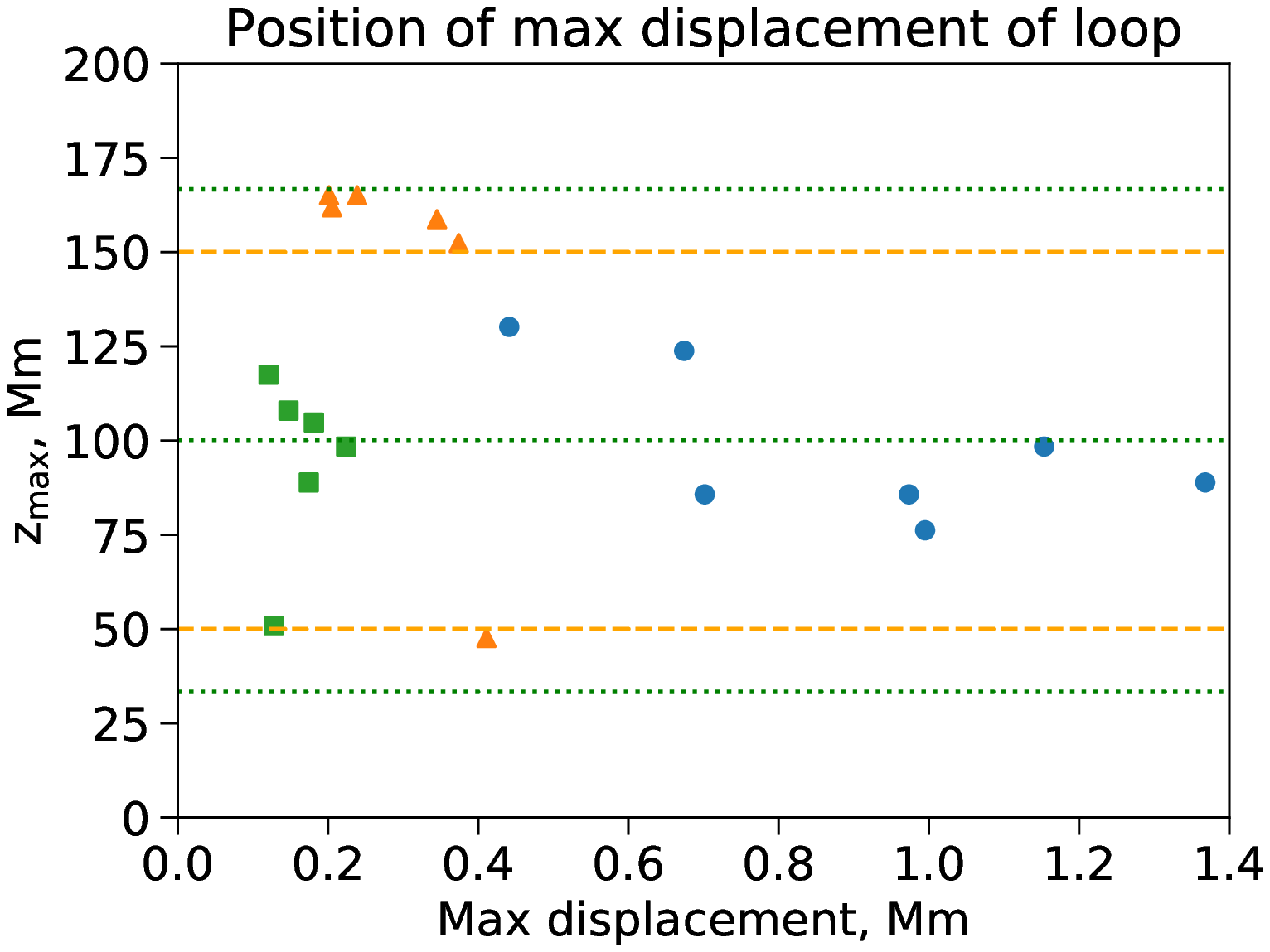}{0.4\textwidth}{(b)}
          }
\caption{Location of the maximum displacement in the loop (Panel \textit{a}), and dependence of the maximum displacement of the loop on the  frequency of the driver exciting the loop footpoint (Panel \textit{b}). Blue points, yellow triangles, and green squares correspond to frequencies in the vicinity of the fundamental mode ($n=1$), second harmonic ($n=2$), and third harmonic ($n=3$), respectively. Yellow dashed and green dotted lines show the expected positions of the standing wave anti-nodes for the second and third harmonics, respectively. \label{fig_resonance_curve}}
\end{figure*}

\begin{figure*}
\gridline{\fig{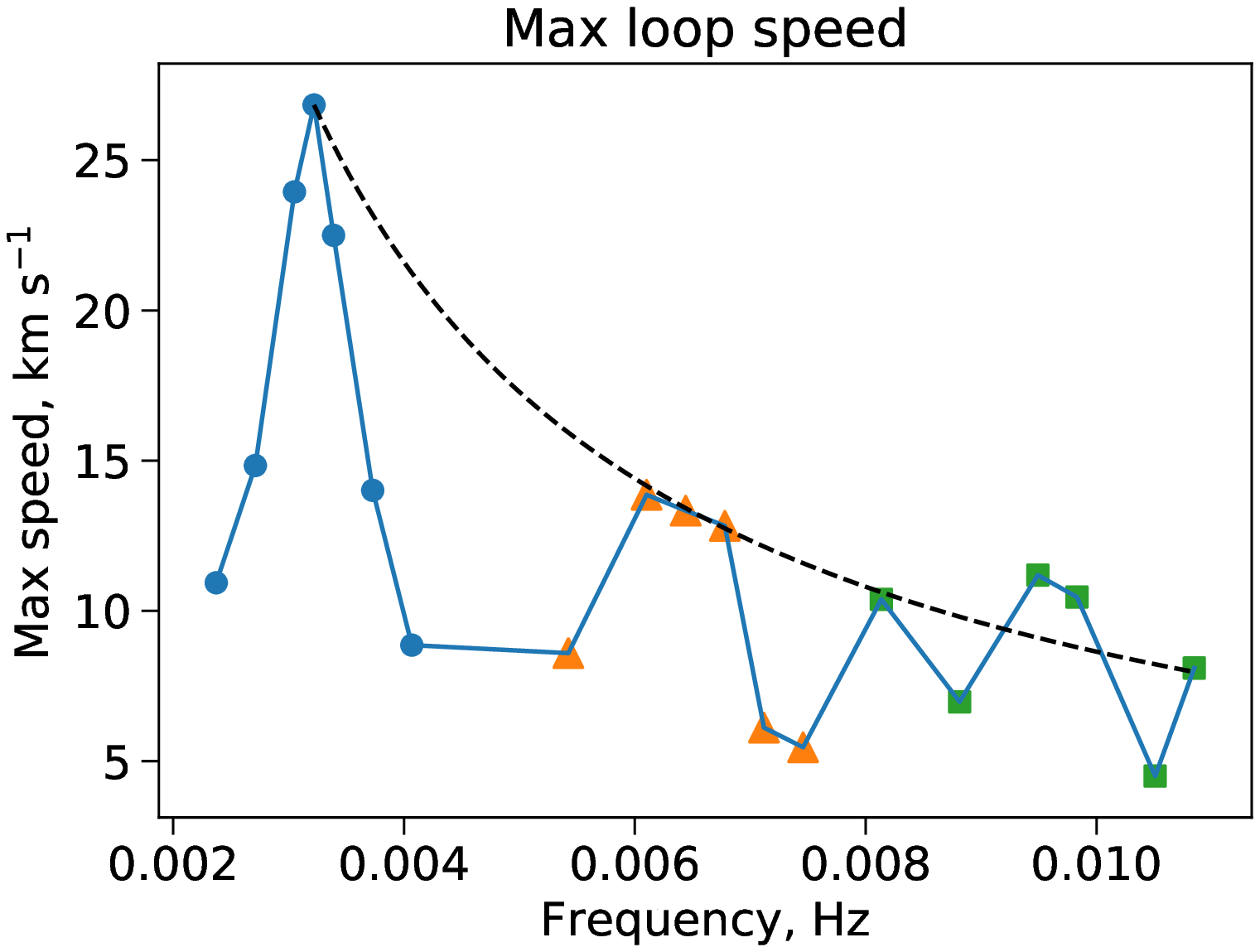}{0.4\textwidth}{(a)}
          \fig{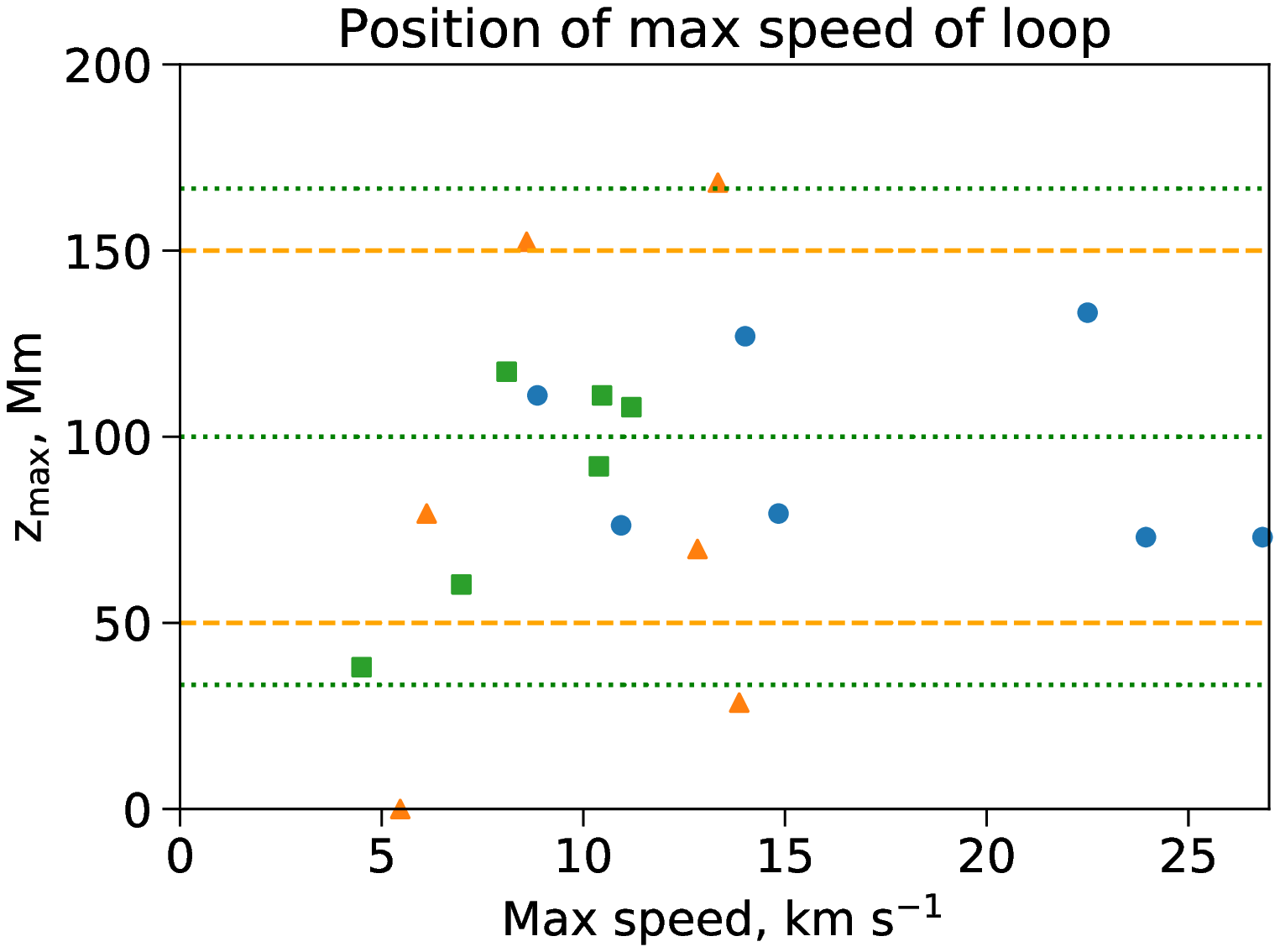}{0.4\textwidth}{(b)}
          }
\caption{Location of the maximum speed of the loop (Panel \textit{a}), and dependence of the maximum speed of the loop on the frequency of the driver exciting the loop footpoint (Panel \textit{b}). The black dashed line represents the expected envelope for a non-stratified loop, the other colour marking is the same as in Fig~\ref{fig_resonance_curve}. \label{fig_resonance_curve2}}
\end{figure*}

In order to demonstrate the response of the loop to its excitation at different frequencies, we have performed 19 simulations. Figure~\ref{fig_resonance_curve}a demonstrates the dependence of the maximum displacement of the loop centre of mass upon the driver frequency. The curve has clear resonant peaks corresponding to periods of $328$~s, $164$~s, $113$~s for the fundamental mode (or first harmonic, with the number of anti-nodes  $n=1$), second harmonic ($n=2$), and third harmonic ($n=3$), respectively. The ratios of the fundamental mode frequency to the second and third harmonic frequencies are $2.0$ and $2.9$, which is quite interesting because very recently, \cite{Duckenfield2018ApJ...854L...5D} detected the first-to-second harmonic frequency ratio for decay-less transverse oscillations in a coronal loop to be $1.4$. These ratios are expected to be different from $n$ due to gravitational stratification of the plasma, in particular, due to the higher kink speed at the loop apex, however, in our case, the plasma temperature inside the loop is relatively high, so the effect is not very significant (cf. the estimates for the harmonic periods with the use of a weight function in Sect.~\ref{sec:model}).
We have presented plots only for one density threshold value used. The behaviour of the resonant curve is similar for the other threshold values, with the resonant peaks being slightly higher for higher threshold values.
Figure~\ref{fig_resonance_curve}b presents the $z$-coordinate of the maximum displacement of the loop centre of mass. The values corresponding to the resonant peaks tend to the expected positions of the standing wave anti-nodes, with this tendency being more pronounced for higher density threshold values. Here, we have taken $\rho_{\mathrm{th}}=0.9$.

However, we should take into account that the displacement of the loop is inversely proportional to the frequency, $Y \propto V_y/f$, where $V_y$ is the speed of the loop centre of mass, $f$ is the oscillation frequency, $Y$ is the displacement. To exclude this effect, we plot in Fig.~\ref{fig_resonance_curve2}a the maximum loop oscillation speeds, differentiating numerically the loop centre of mass displacement data. The resonant frequencies are slightly shifted, so we have for the fundamental mode a period of $310$~s (cf. the value predicted analytically, see Sect.~\ref{sec:model}), and the first-to-second harmonic frequency ratio of $1.9$. However, those frequency shifts depend on the density threshold value chosen ( $\rho_{\mathrm{th}}=0.7$ was used for these plots). Some error in loop speed values is contributed by differentiating numerically, which is confirmed by comparison of the loop centre of mass displacement calculated by integrating the loop speed to the original loop displacement values. In particular, the obtained information is less reliable, at least, for the third harmonic ($n=3$), whose period is shifted to $105$~s. The same numerical effects might cause the significant scattering of points in Fig.~\ref{fig_resonance_curve2}b.

On the other hand, oscillation speed values are determined by re-distribution of the wave energy delivered by the footpoint driver over the period among several oscillation anti-nodes. So, for a uniform (without stratification) loop, one can expect the $1/f$ envelope for the resonant frequencies. The dashed line in Fig.~\ref{fig_resonance_curve2}a shows this envelope. One more factor affecting the amplitude of resonant harmonics is the gravitational stratification of the plasma in the loop. Due to stratification and the higher kink speed, oscillations near the loop apex are stronger (see, for instance, numerical and analytical results by \citealp{Verth-etal2007} and \citealp{ErdelyiVerth2007}). So, the higher values in the vicinity of the third harmonic ($n=3$) in Fig.~\ref{fig_resonance_curve2}a could be due to the gravitational stratification effect because of the location of its maximum near the loop apex.

The plots presented in Fig.~\ref{fig_resonance_curve}a and Fig.~\ref{fig_resonance_curve2}a show the response of the loop to driving it at different frequencies. However, these plots are not eigenspectra of the system because loop oscillation signals contain different harmonics, so Fourier decomposition is required to obtain pure spectra. Although such information could be of some importance, we do not obtain them, analysing only displacements and speeds of the loop.

\begin{figure*}
\gridline{\fig{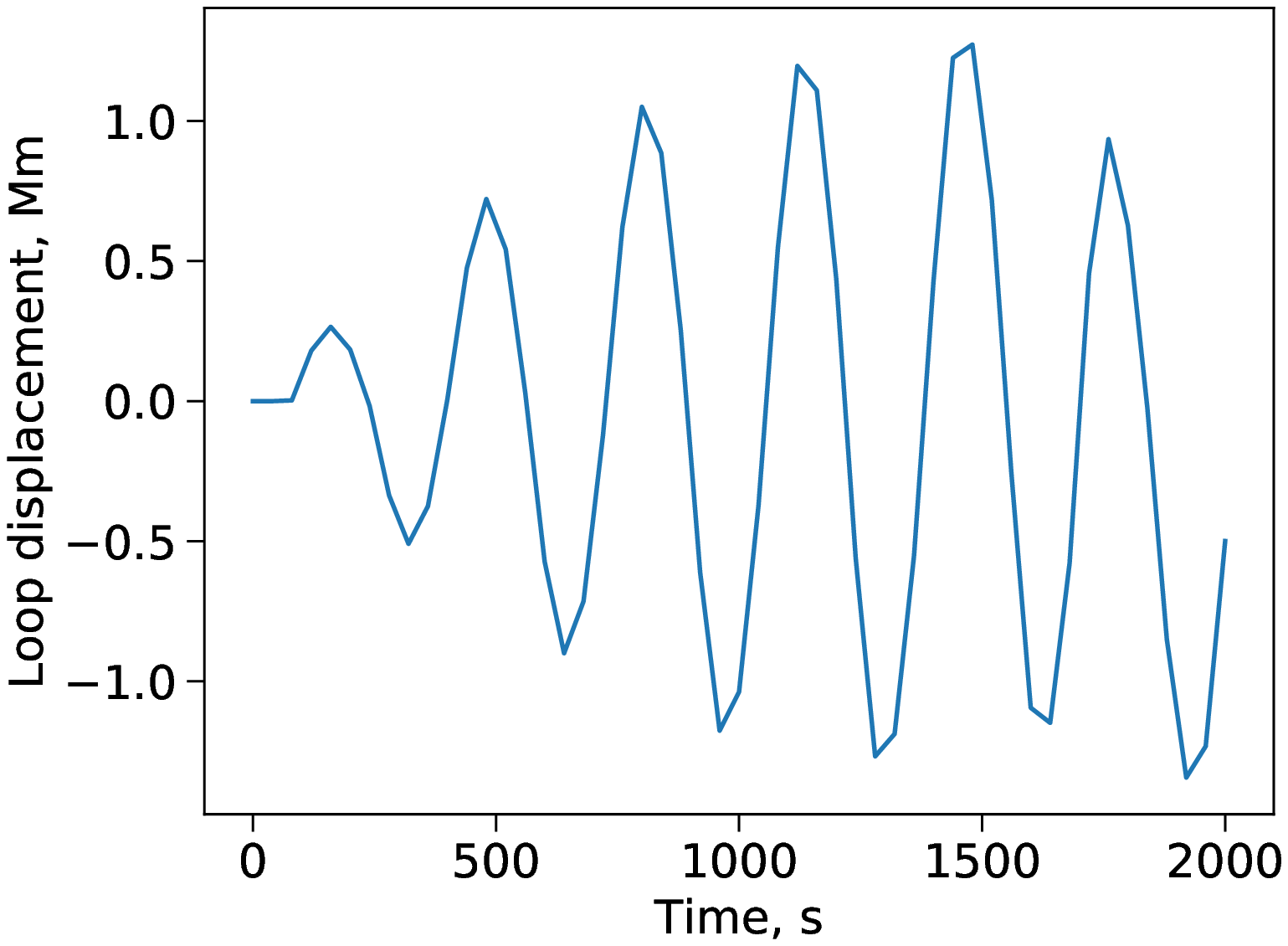}{0.3\textwidth}{(a)}
          \fig{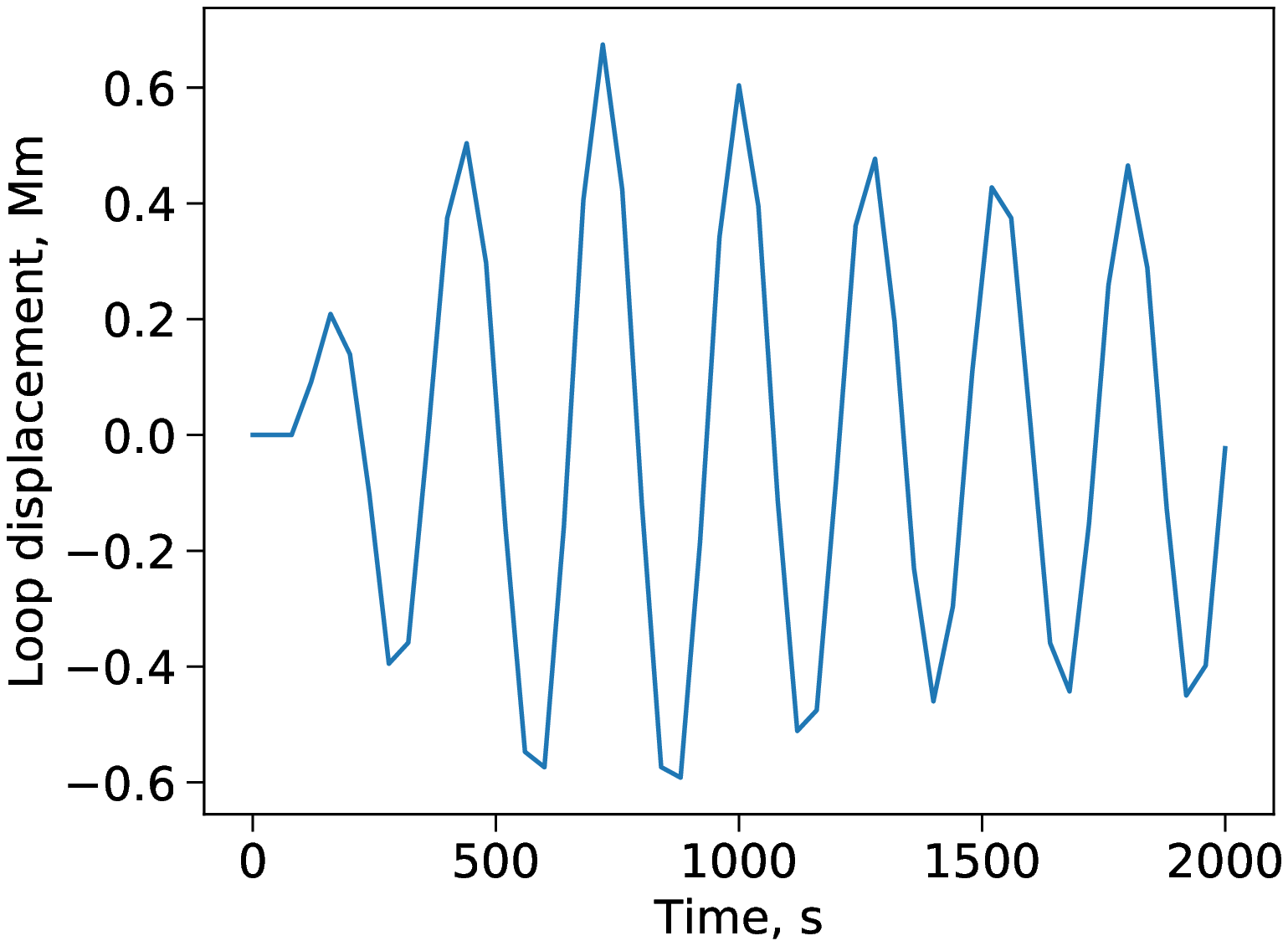}{0.3\textwidth}{(b)}
          \fig{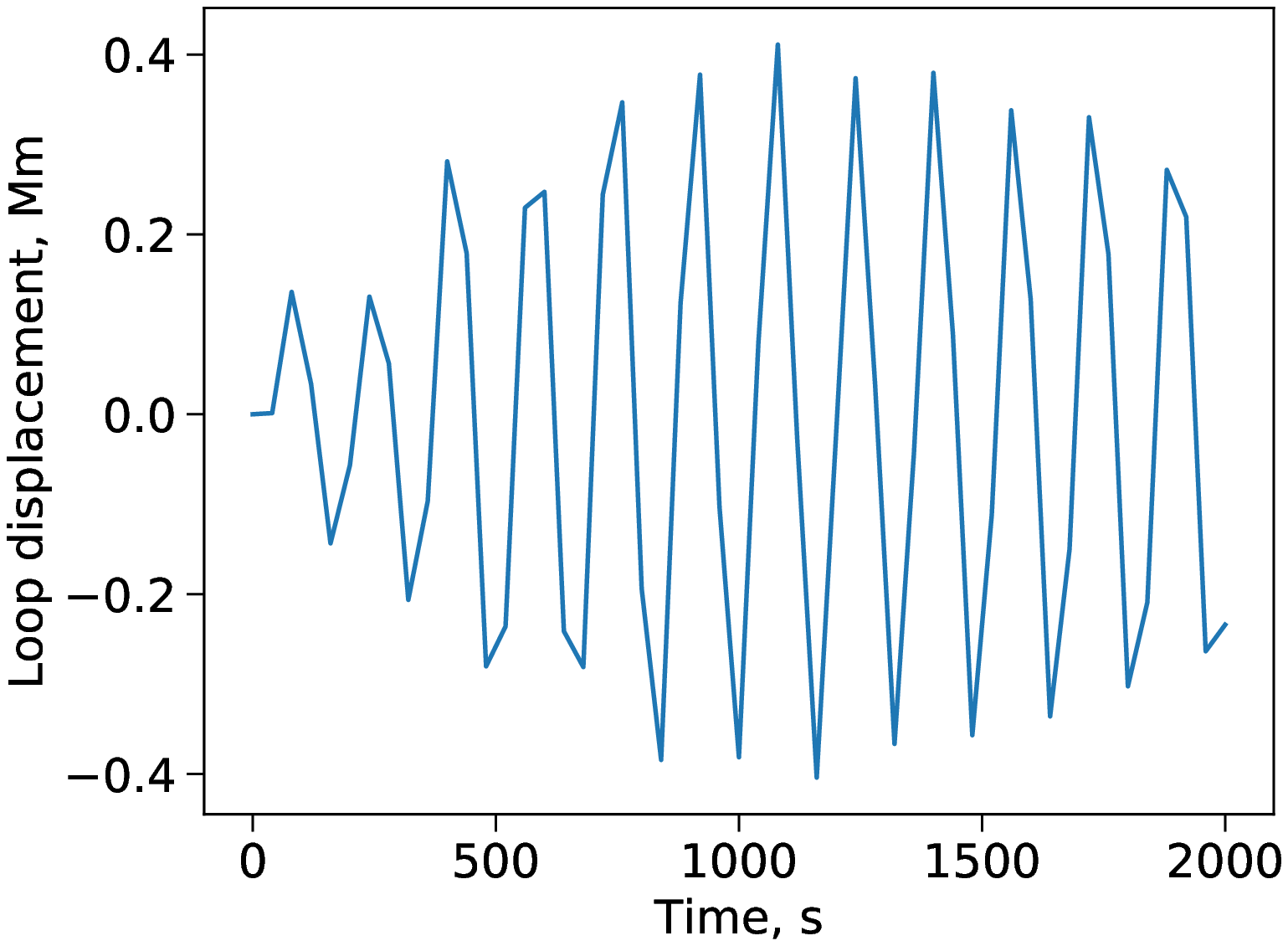}{0.3\textwidth}{(c)}
          }
\gridline{\fig{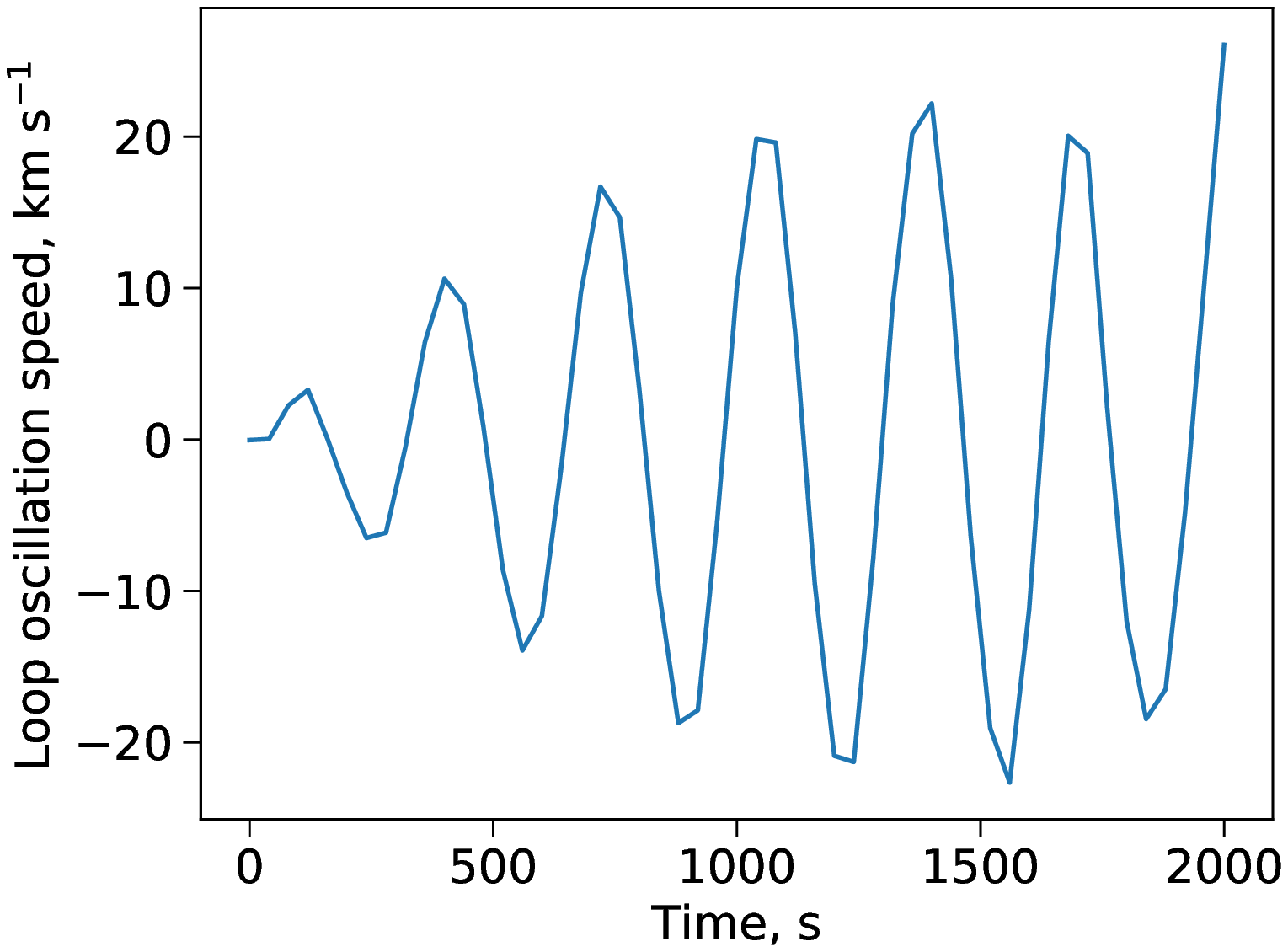}{0.3\textwidth}{}
          \fig{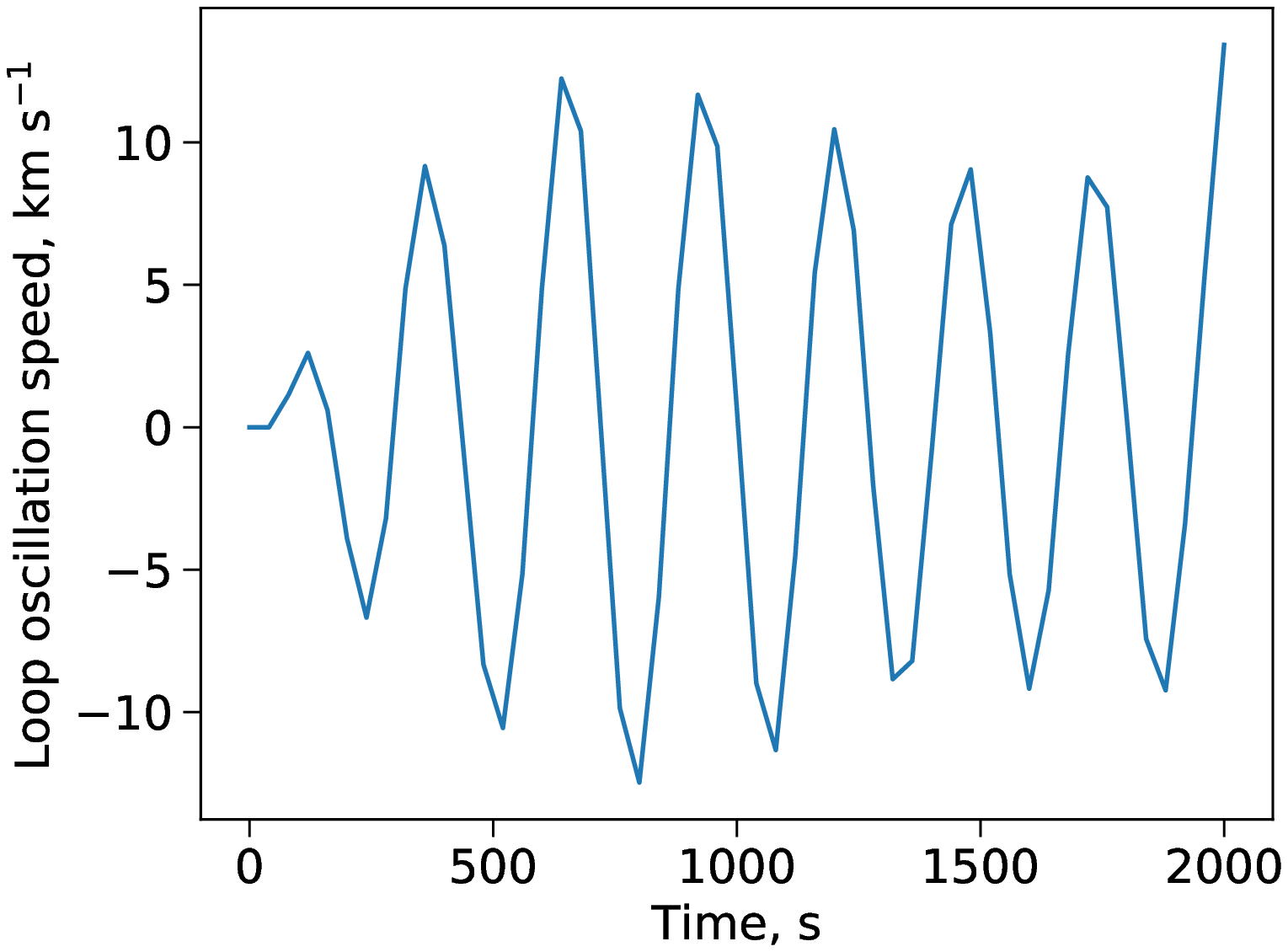}{0.3\textwidth}{}
          \fig{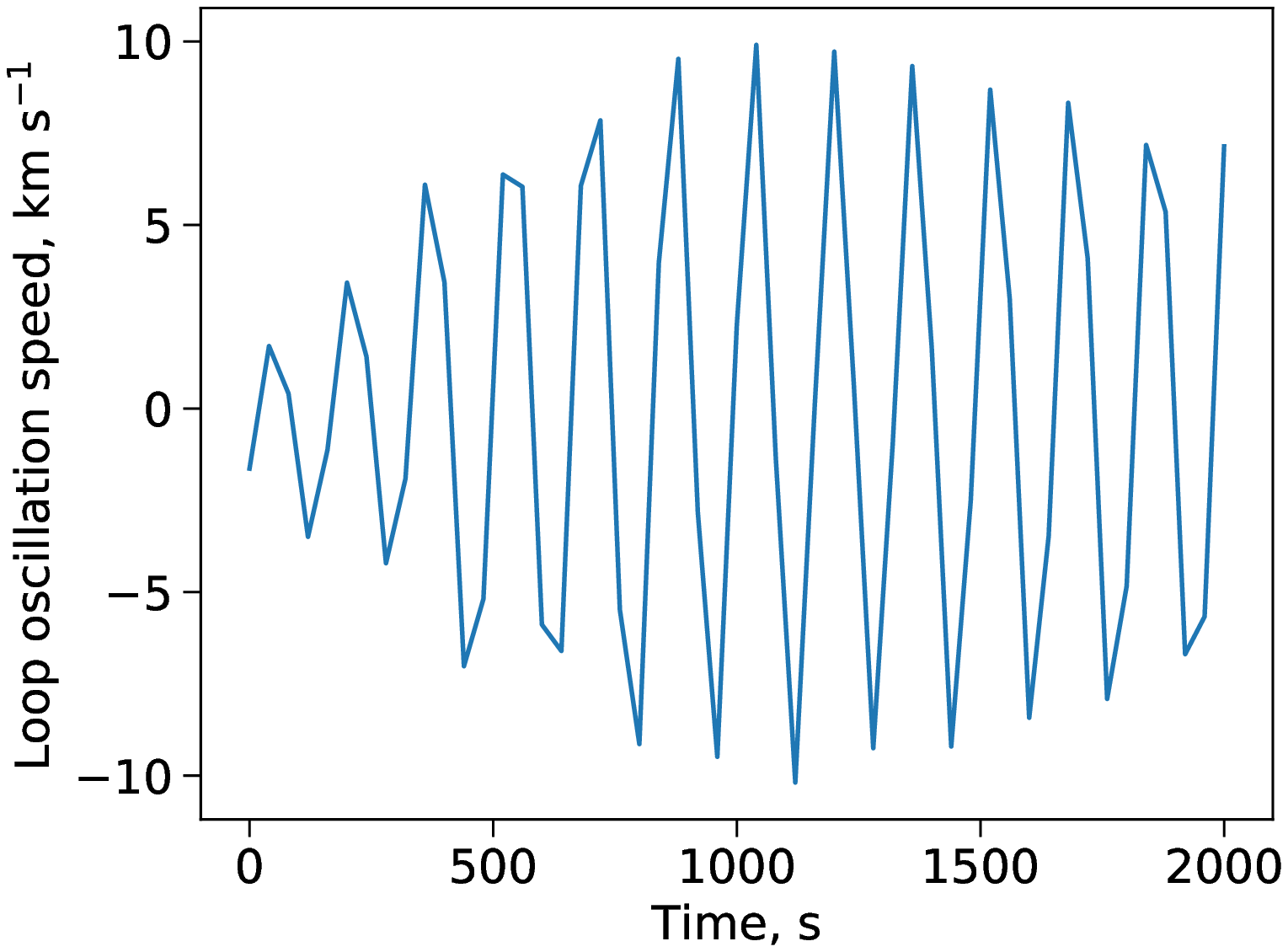}{0.3\textwidth}{}
          }
\caption{Displacement (top panels) and oscillation speed (bottom panels) of the loop centre of mass for different driver periods and heights corresponding to: (a) the fundamental mode ($n=1$) and $z \approx 100$~Mm, (b) an intermediate value between the first and second harmonics and $z \approx 125$~Mm, (c) the second harmonic ($n=2$) and $z \approx 100$~Mm. \label{fig_displacement_vs_time}}
\end{figure*}

Figure~\ref{fig_displacement_vs_time} shows the displacement of the loop centre of mass and its speed at heights corresponding roughly to the maximum loop displacement for three different periods of the driver: the fundamental mode ($P_1 \approx 328$~s), an intermediate value between the first and second harmonics ($P_{12} \approx 268$~s), and the second harmonic ($P_2 \approx 164$~s). The threshold value used is $\rho_{\mathrm{th}}=0.9$. In the resonant cases (panels \textit{a} and \textit{c}), the amplitude of loop oscillations grows until the turbulence has developed near the loop boundary (see Fig.~\ref{fig_vorticity} below). After that, oscillations decay, which appears to be due to the contribution of KHI eddies developed (see, e.g., \citealp{Karampelas2017}). The analysis of the loop displacement for different threshold values, $\rho_{\mathrm{th}}$, shows a slight enhancement of the oscillation amplitude for higher threshold values. 


\begin{figure*}
\plotone{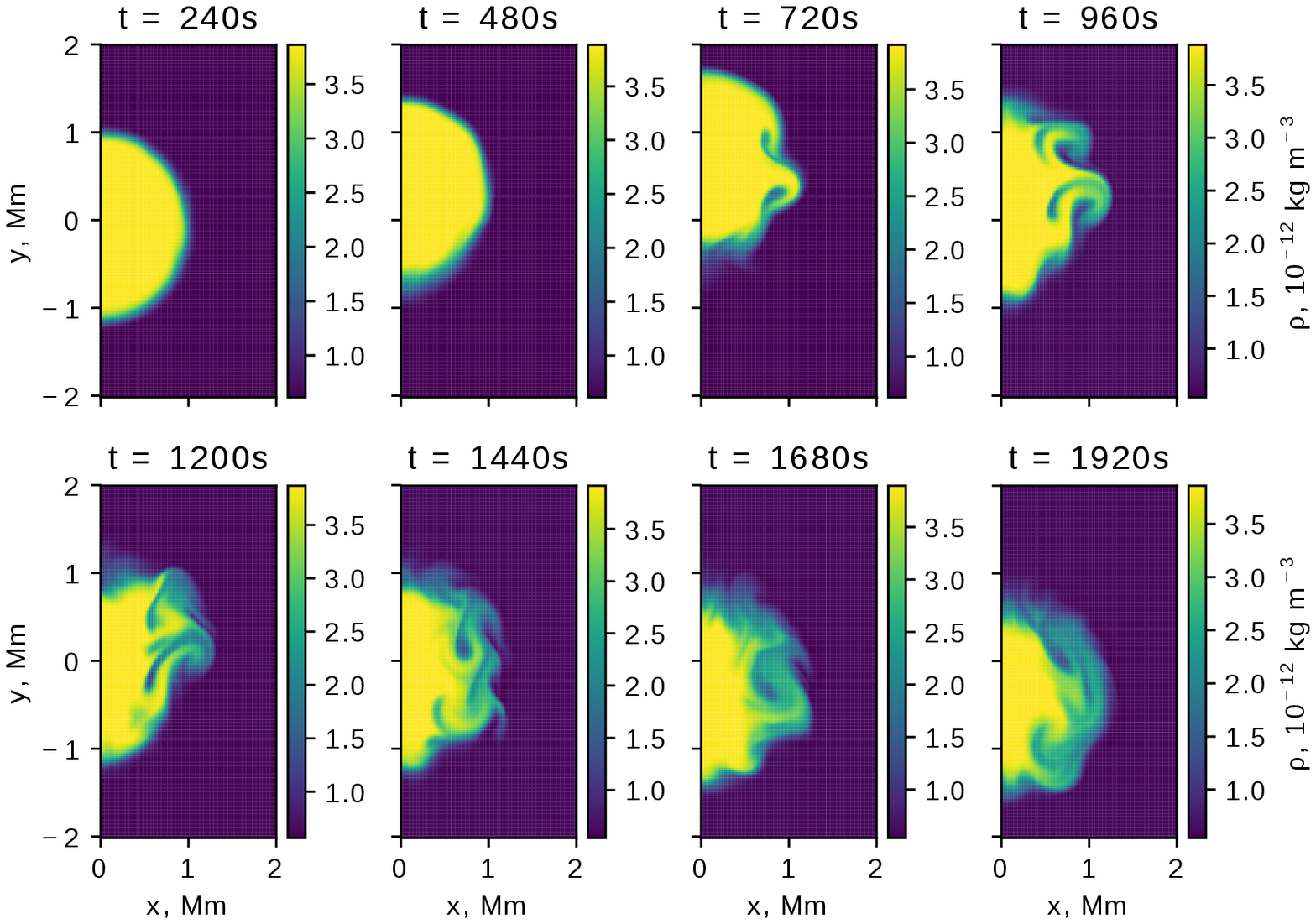}
\caption{Development of KHI in the intermediate case of the driver period between the first and second harmonics. The cross-section of the loop is taken at $z \approx 123$~Mm, which corresponds to the position of maximum loop displacement for that period. \label{fig_KHI_development}}
\end{figure*}

\begin{figure}
   \centering
   \includegraphics[width=0.4\hsize, trim={5cm 0 0 0}, clip]{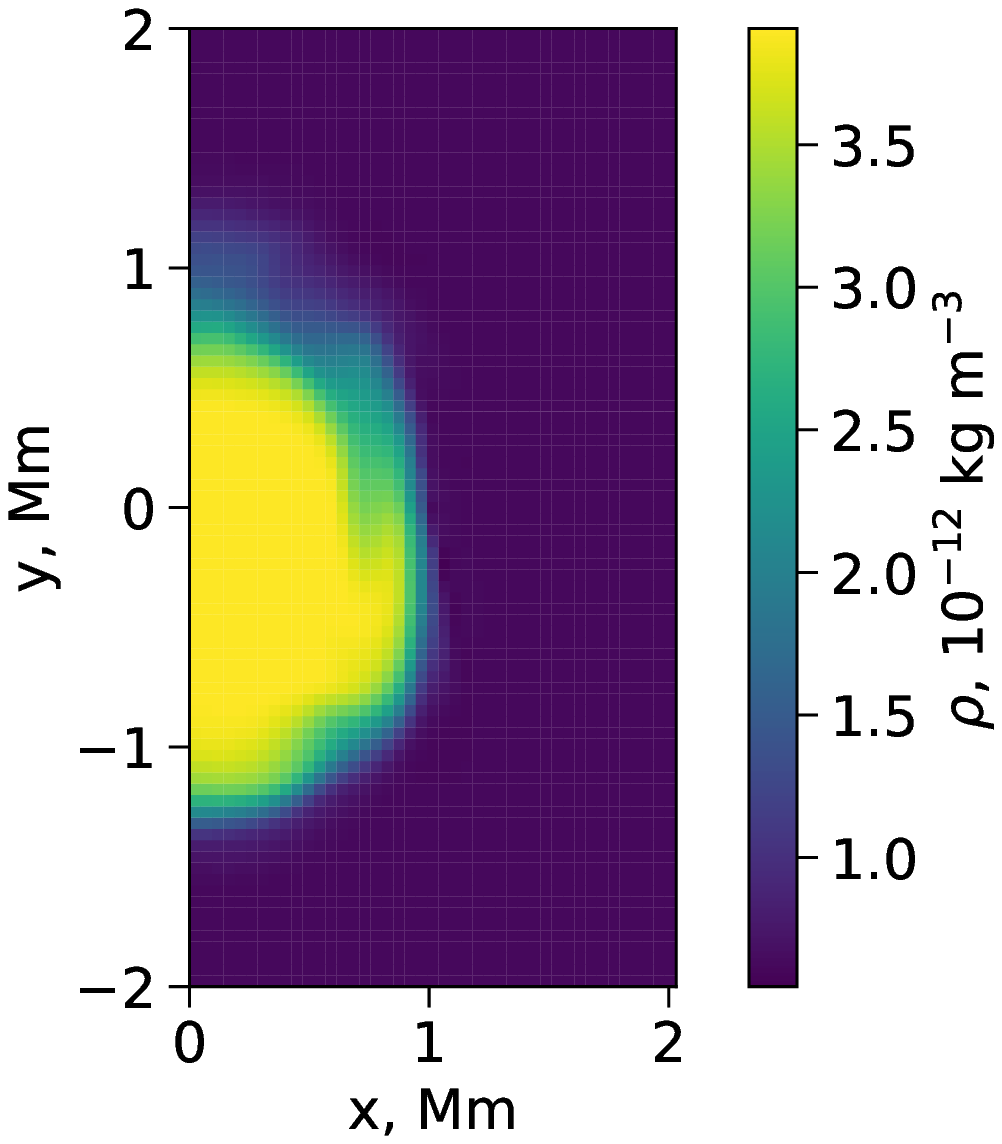}
   \hspace*{0.5cm}
   \includegraphics[width=0.4\hsize, trim={5cm 0 0 0}, clip]{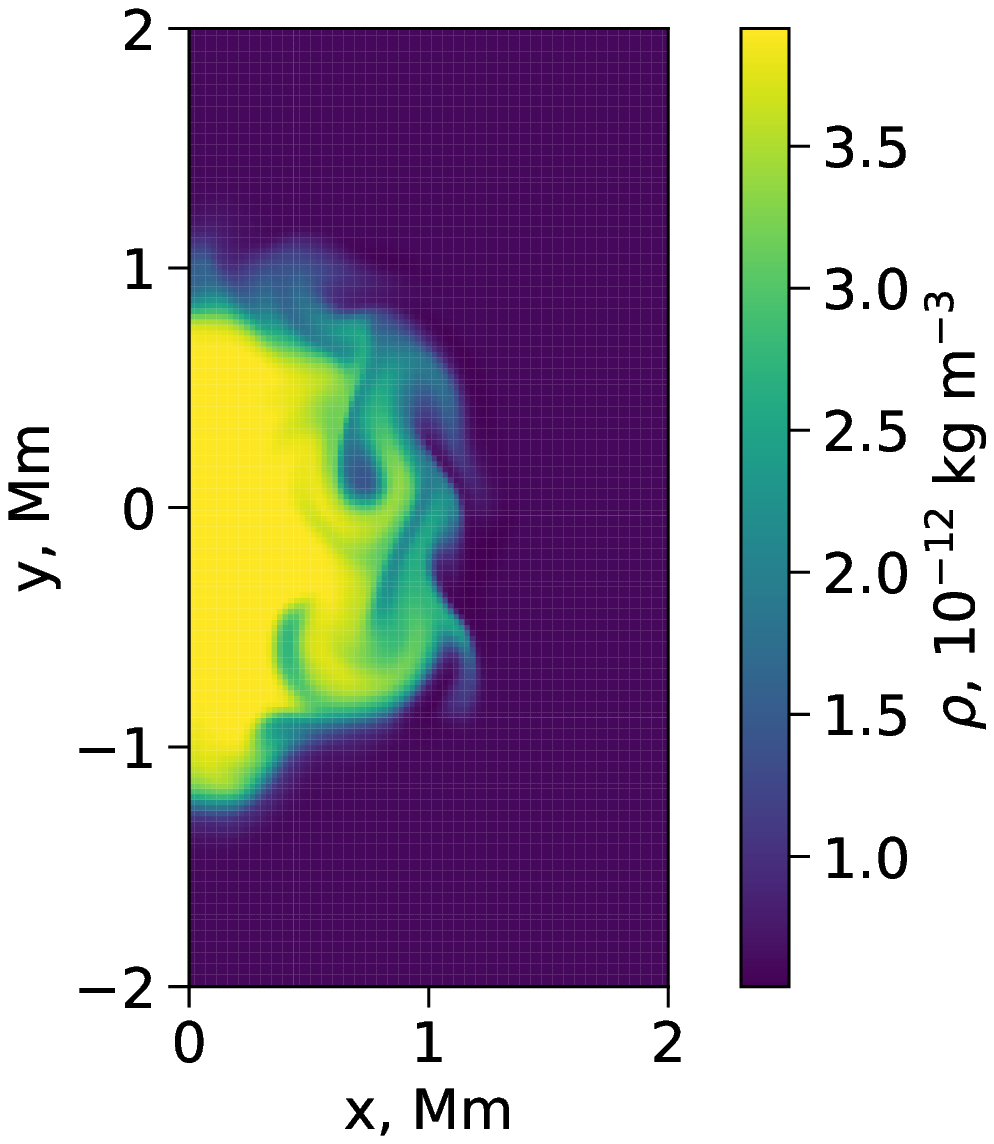}
   \caption{Density snapshots at $t=1440$~s from simulations with the single (left panel) and double (right panel) resolution. The position of the cross-section and value of the driver period is the same as in Fig.~\ref{fig_KHI_development}.}
\label{fig_KHI_comparison}%
\end{figure}

In agreement with previous studies, due to shear flows, KHI develops near the loop boundary, significantly distorting its shape. Figure~\ref{fig_KHI_development} shows the development of KHI in an intermediate case of the driver period between the first and second harmonics ($P_{12} \approx 268$~s). The amplitude of the loop displacement is lower than that in the resonant cases. The presented plots show the simulation performed with the double resolution. The comparison to results of the lower-resolution simulations demonstrates that numerical diffusion due to the finite cell size significantly inhibits the appearance of the fine structure of KHI eddies, as seen in Fig.~\ref{fig_KHI_comparison}.

In Fig~\ref{fig_vorticity}, we analyse the magnitude of the $z$-component of the plasma flow vorticity, $\left| \left( {\bf \nabla \times v} \right)_z \right|$, plotting its absolute values averaged over the loop cross-section and its vicinity ($r<2.5$~Mm) for each $z$ as a function of time. Positions of the maximum values of the mean vorticity agree well with those of the loop oscillation anti-nodes, which confirms that KHI develops more intensively where the amplitude of the loop displacement is higher. Similarly, KHI eddies are more powerful in the cases of resonant frequencies of the driver, although the magnitude of the $z$-vorticity for the third harmonic ($n=3$) is higher but comparable to that in the intermediate case with the period between the first and second harmonics ($P_{12} \approx 245$~s). Note also that the turbulent state with a relatively high value of the vorticity is reached at different moments for different periods. However, the time required corresponds in each case to 5--6 driver periods after starting the simulations.

\begin{figure*}
\gridline{\fig{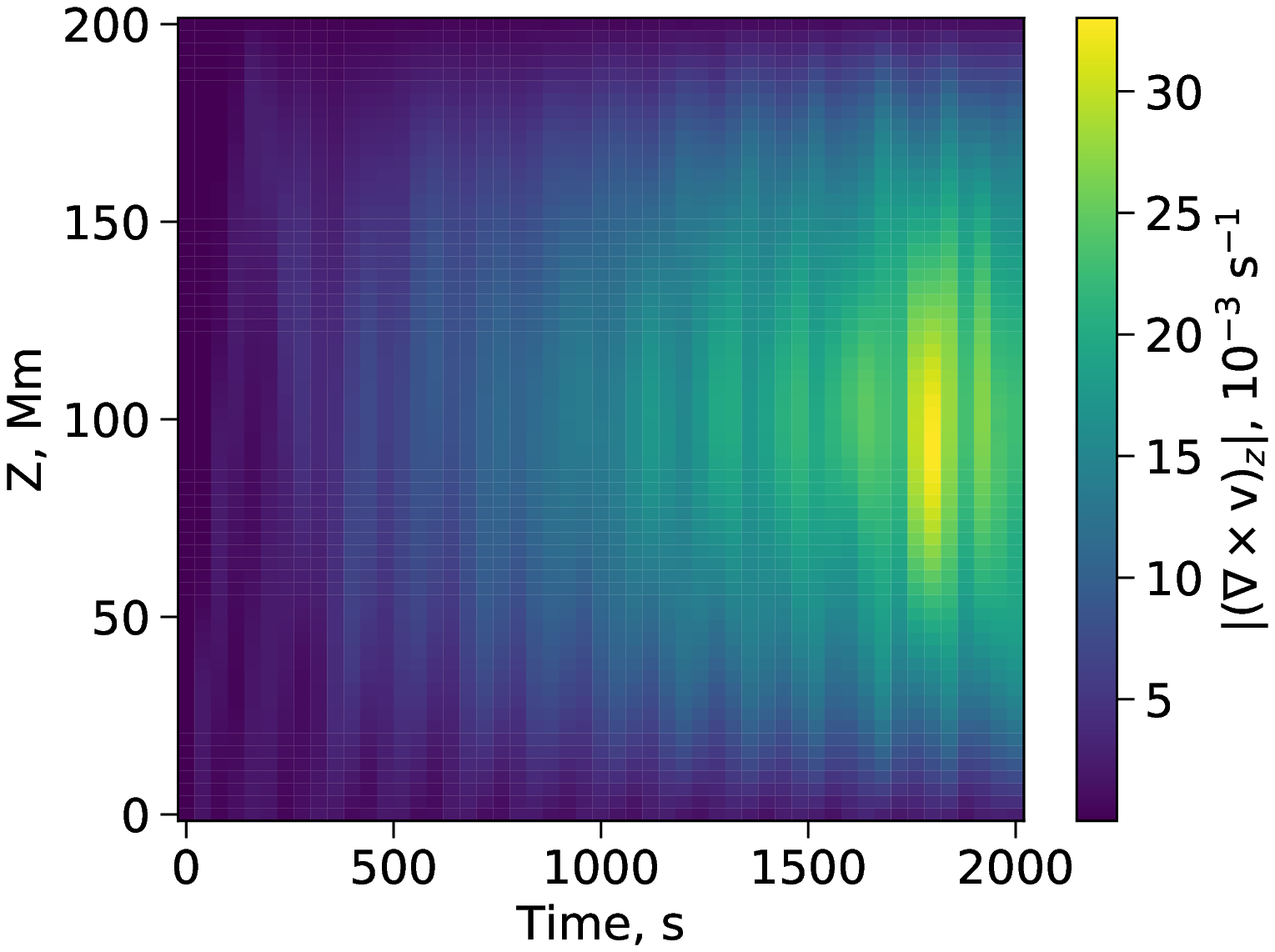}{0.4\textwidth}{(a)}
          \fig{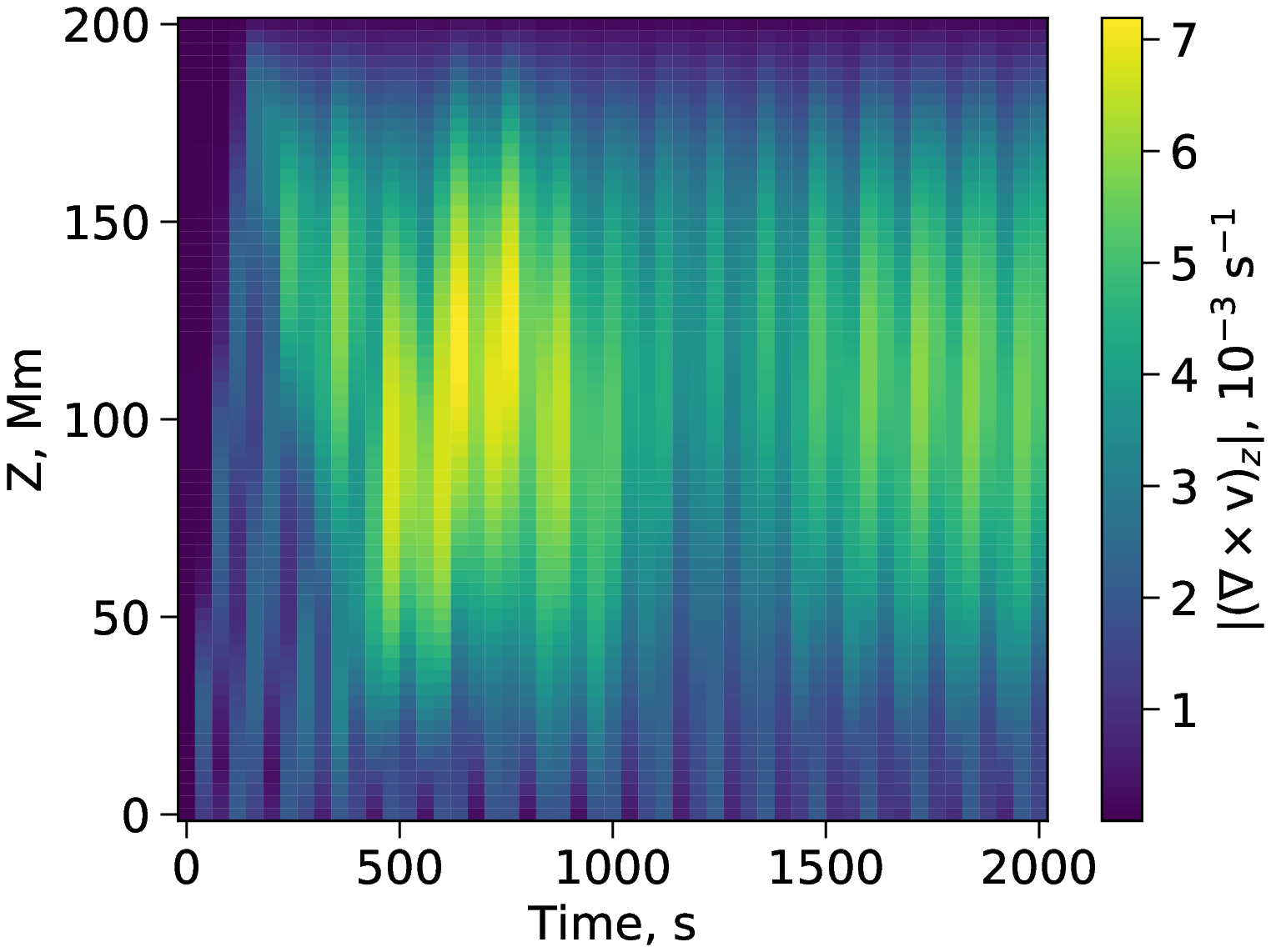}{0.4\textwidth}{(b)}
          }
\gridline{\fig{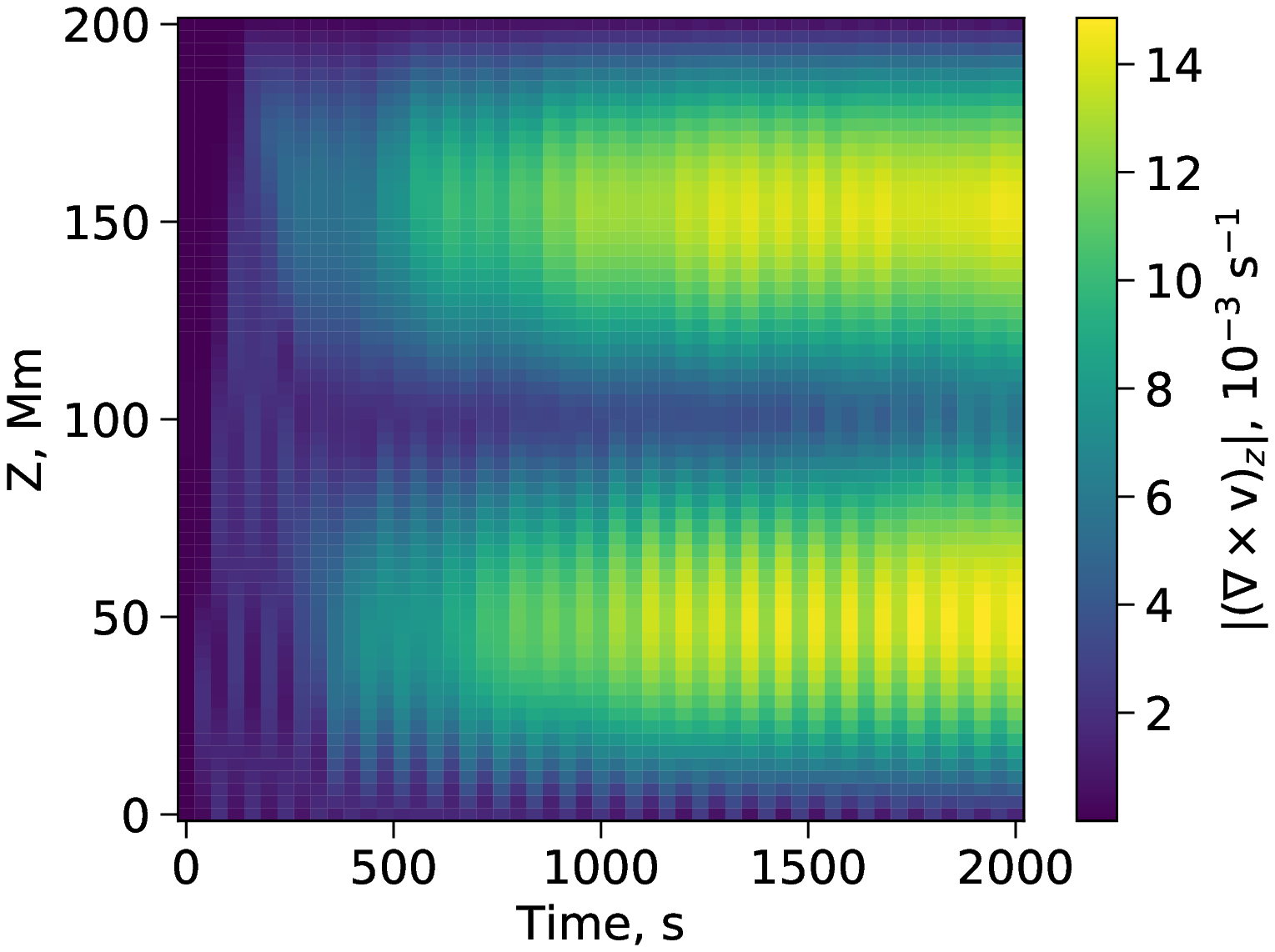}{0.4\textwidth}{(c)}
          \fig{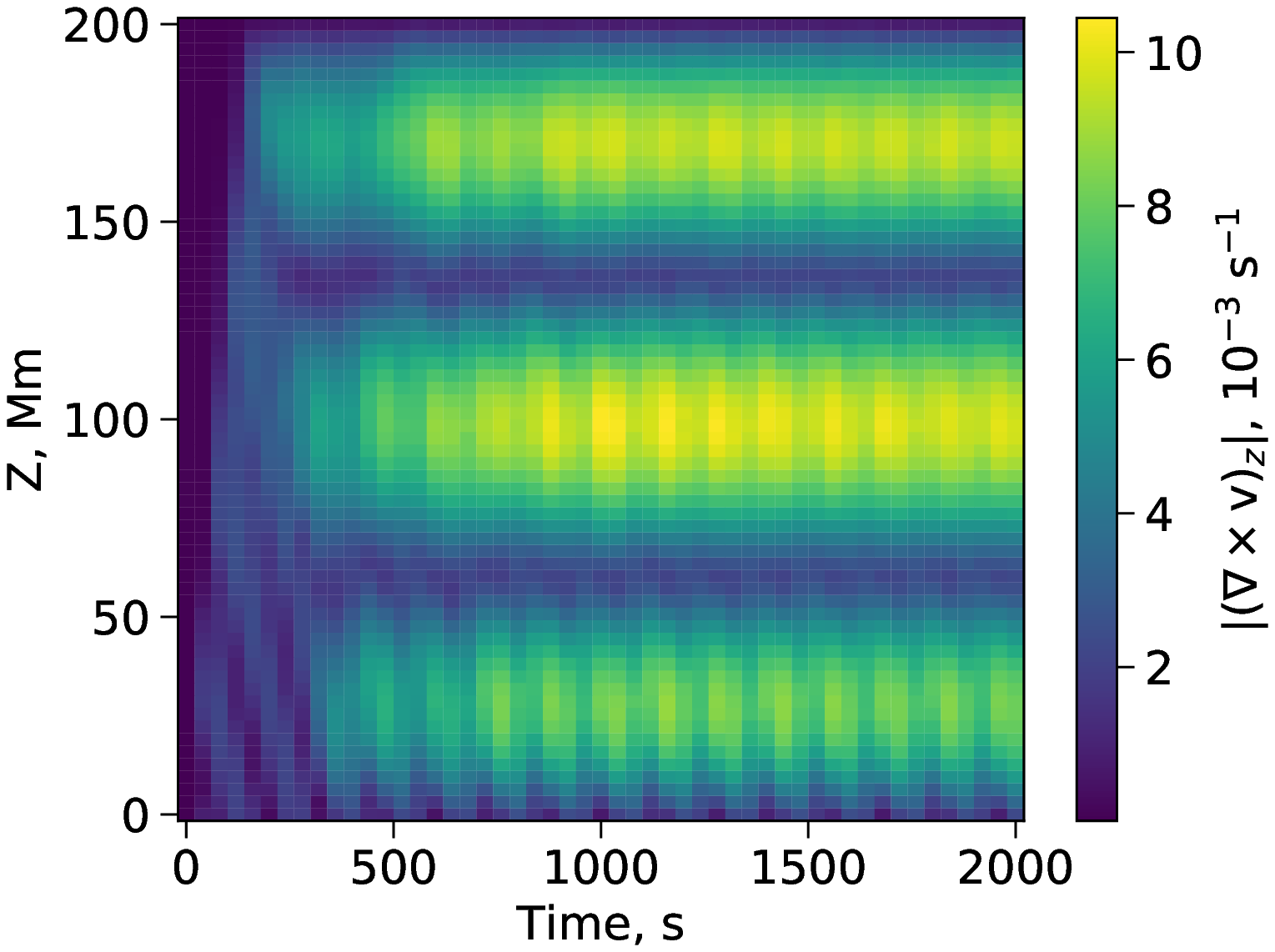}{0.4\textwidth}{(d)}
          }
\caption{Behaviour of the $xy$-averaged magnitude of the $z$-component of the flow vorticity for different periods corresponding to: (a) the fundamental mode ($n=1$), (b) an intermediate value between the first and second harmonics, (c) the second harmonic ($n=2$), (d) the third harmonic ($n=3$). \label{fig_vorticity}}
\end{figure*}


\begin{figure*}
\gridline{\fig{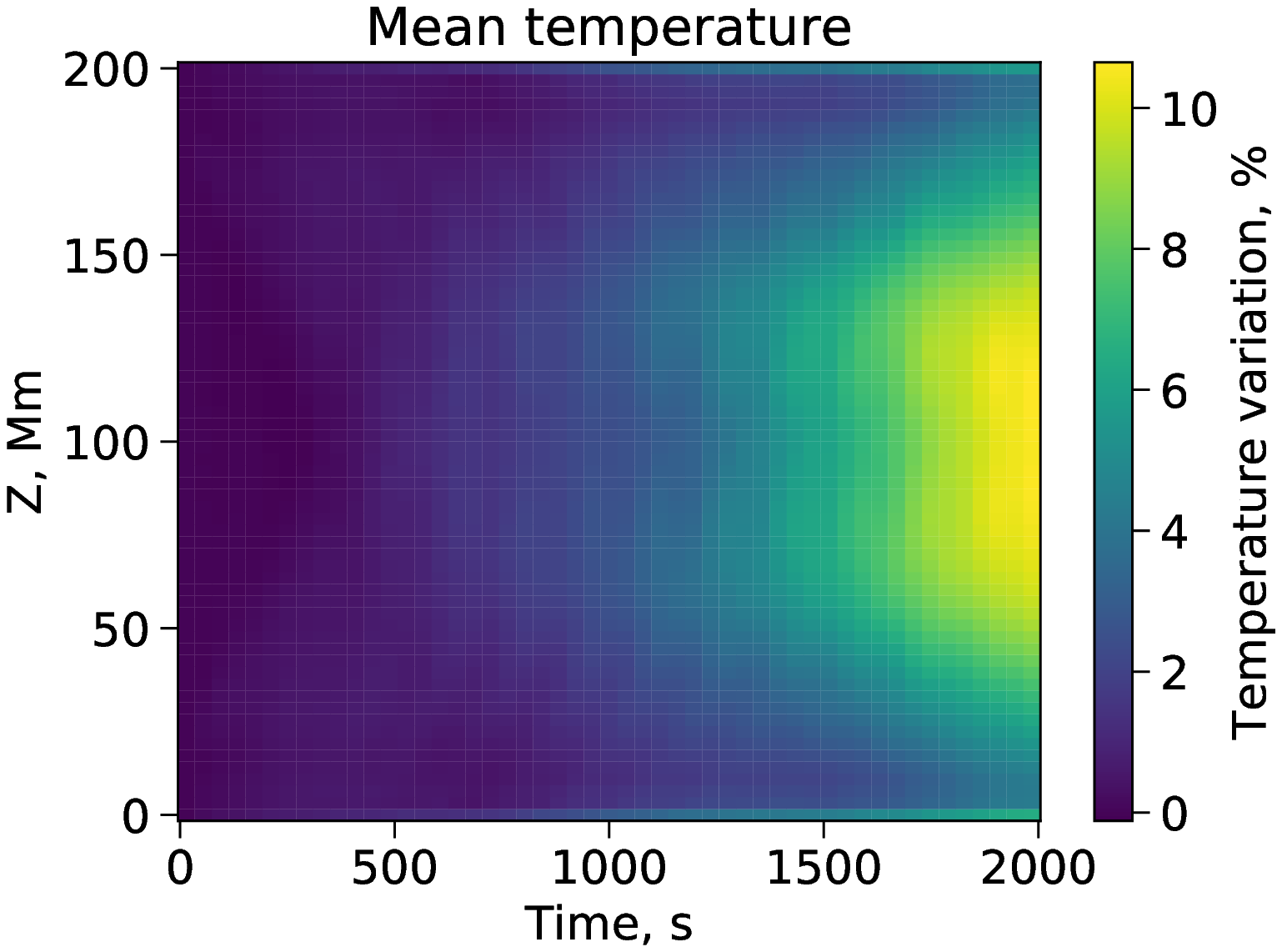}{0.4\textwidth}{(a)}
          \fig{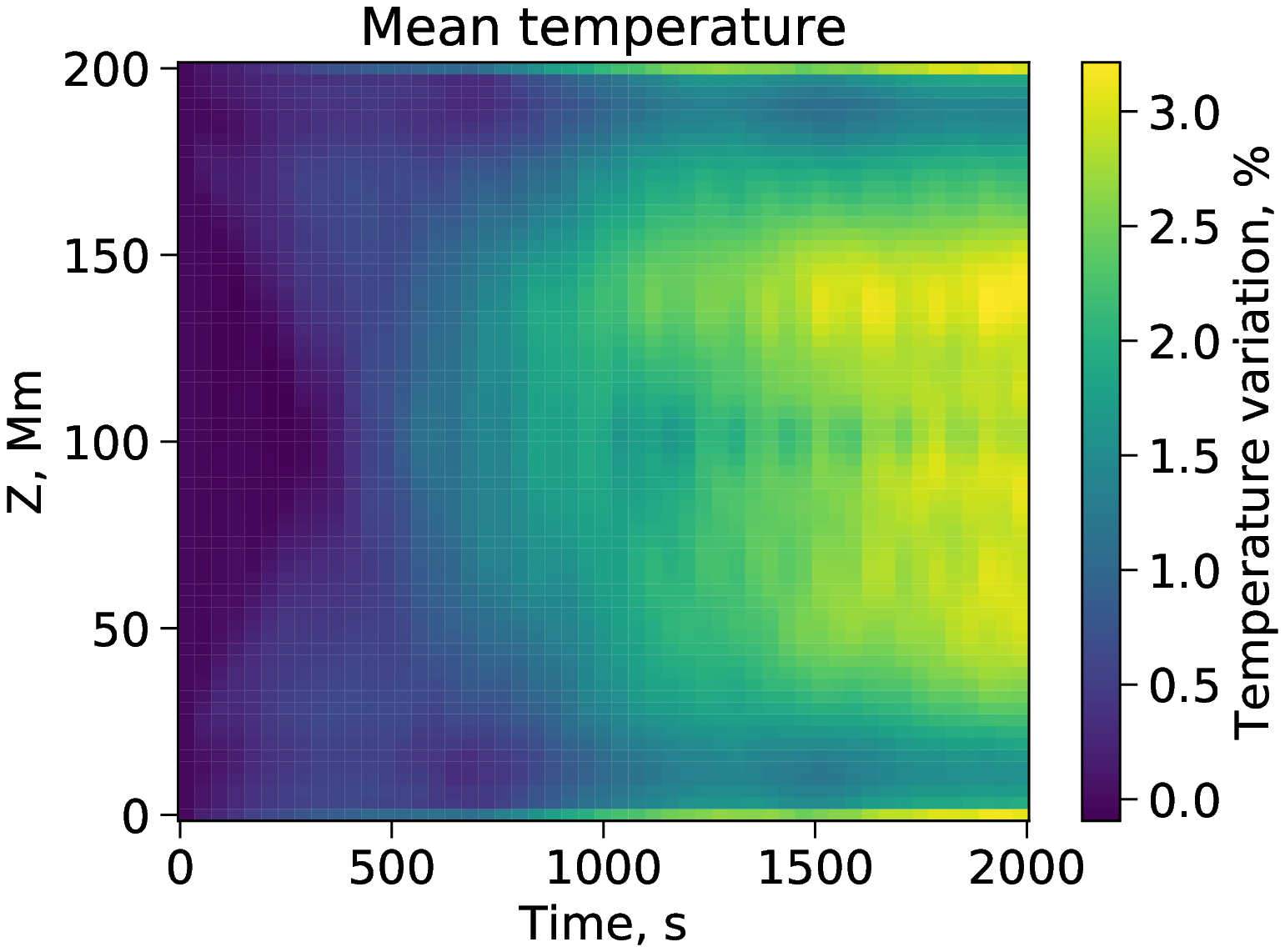}{0.4\textwidth}{(b)}
          }
\gridline{\fig{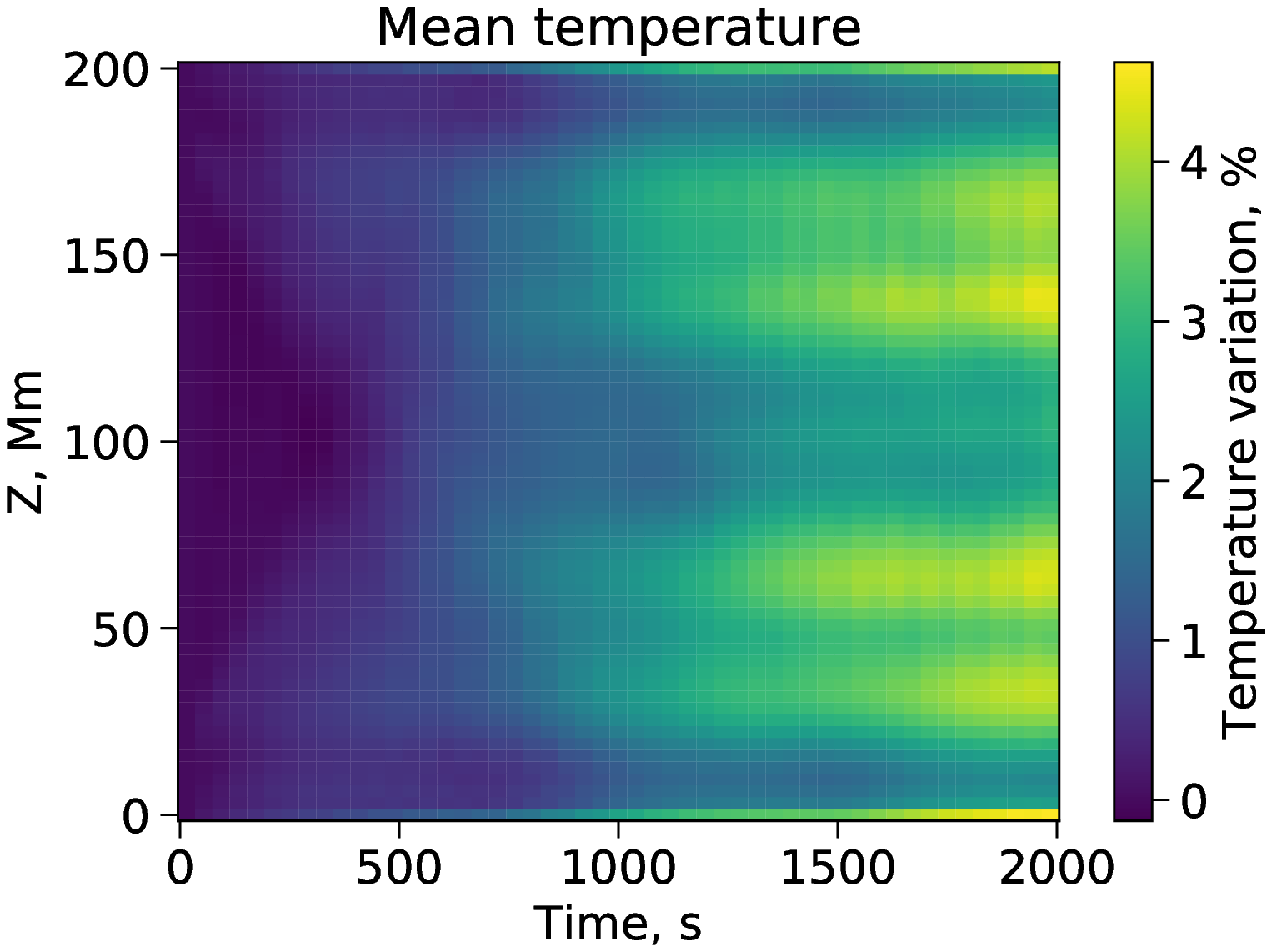}{0.4\textwidth}{(c)}
          \fig{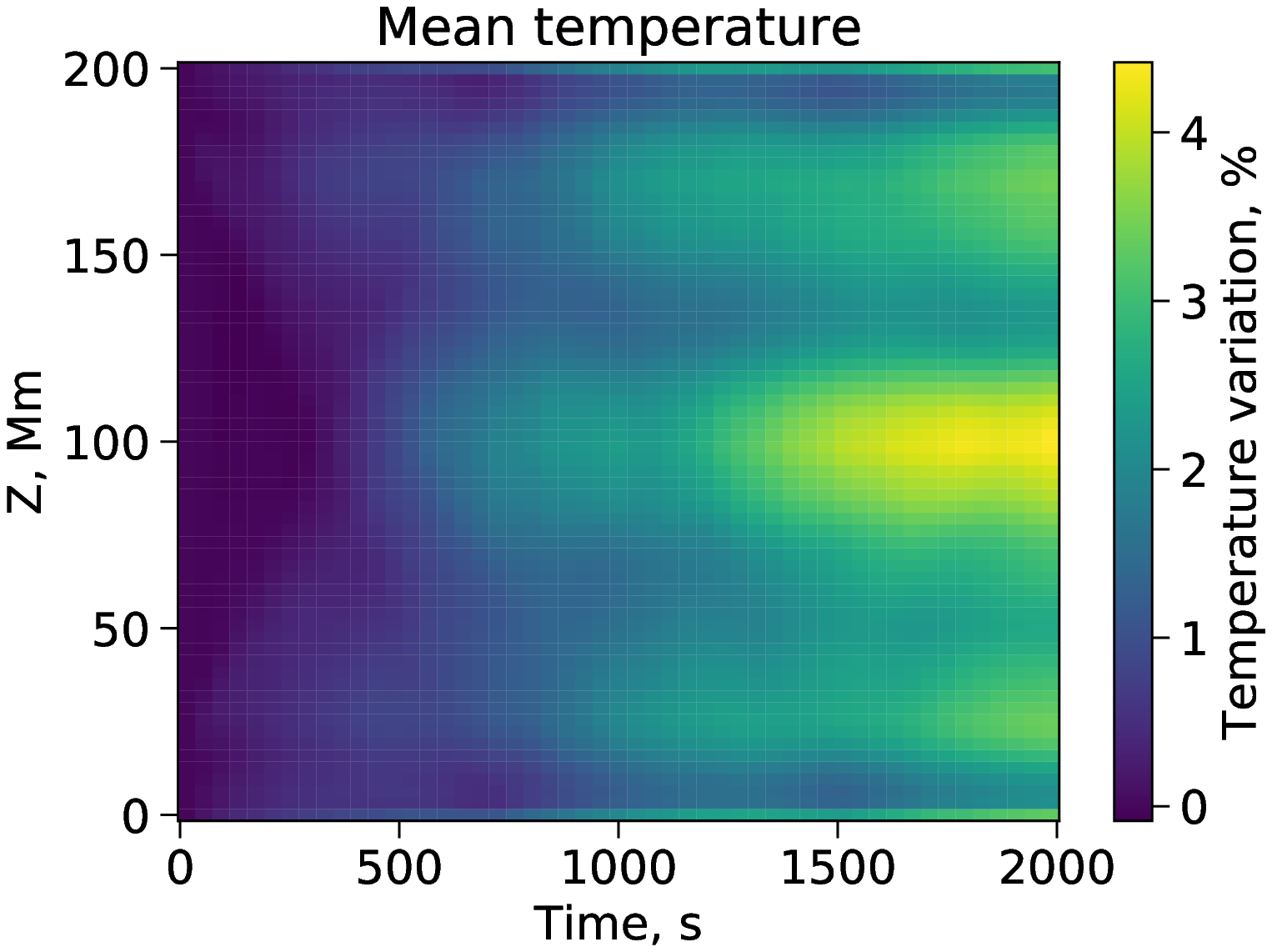}{0.4\textwidth}{(d)}
          }
\caption{Behaviour of the plasma temperature averaged at the $r<2.5$~Mm region for different periods corresponding to: (a) the fundamental mode ($n=1$), (b) an intermediate value between the first and second harmonics, (c) the second harmonic ($n=2$), (d) the third harmonic ($n=3$). \label{fig_mean_temperature}}
\end{figure*}

Figure~\ref{fig_mean_temperature} shows the variation of the plasma temperature averaged for each $z$ over the region including both the loop and its environment. At the heights corresponding to maximum displacements of the loop (and the $z$-component of the vorticity), we see an increase in the plasma temperature. As expected, the increase is higher for resonant periods and lower for the intermediate one ($P_{12} \approx 268$~s).

The obtained temperature enhancement appears to be due to the KHI mixing of plasmas of different temperatures. However, simple redistribution of different fluid parcels in space due to mixing cannot affect the mean value, so adiabatic fluctuations inside them due to pressure and density fluctuations produced by the development of KHI come into play. \citet{Karampelas2017} showed that adiabatic fluctuations in plasmas of equal temperatures inside and outside the loop without gravity affect very slightly the mean temperature value. In the case of different temperatures and densities of the loop and its environment, KHI could cause different temperature fluctuations inside and outside the loop, leading to the temperature enhancement inside the domain considered.

The plasma temperature enhancement found in our simulations together with results of previous studies show an interesting tendency. To discuss it, we recall the results by \citet{Karampelas2017} for loop oscillations in the absence of gravity, and very recent results by \citet{Karampelas2019}, in which the gravity was taken into account.
In both studies, the authors considered the averaged temperature for the greater region including both the loop and its environment.
In the case of equal temperatures of the loop and its environment, \citet{Karampelas2017} found that only minor heating of the plasma occurred at the loop apex, which was completely hidden by the effects of Ohmic heating near the footpoints.
In the case of a colder loop and hotter environment, they found cooling of the plasma. The obtained cooling effect disappeared towards the loop footpoints, which implied the contribution of KHI mixing. Unlike the equal temperature case, the heating due to Ohmic dissipation was dominated by effects of plasma mixing.
In \citet{Karampelas2019}, the same problem of a colder loop in a hotter environment was considered in a gravitationally stratified plasma. Although weak heating was found both at the apex and footpoints, the effect at the loop apex appeared only at the end of their extended simulations, and before that, the plasma cooling occurred there.
Finally, in this study for a gravitationally stratified hotter loop in a colder environment, we have obtained a more significant averaged temperature enhancement at the loop apex where KHI eddies are the strongest.
Thus, it is possible to conclude that due to KHI, heating has been found for a hotter loop, whereas cooling has been found for a colder loop, followed by weak heating in the case of a gravitationally stratified loop.

The simulations performed with the double resolution have shown only several per cent difference for the maximum loop displacement, while the flow vorticity and magnitude of the averaged temperature variation differ by up to two times, with their values being larger for the simulations with the higher resolution. This is in agreement with the interpretation in terms of plasma mixing and could be due the turbulent nature of the process under study. The finer simulations we perform, the more and more the fine structures of KHI eddies affect the plasma temperature. The same result was obtained by \citet{Antolin2015}. Therefore, it seems that because of the self-similarity of the turbulence, one could never reach the convergence of simulations. However, the natural limit for that is associated with real dissipation in the plasma and determined by coronal values of Lundquist and Reynolds numbers, which cannot be incorporated in simulations. We will address the issue how the temperature enhancement scales with respect to the resolution used in the subsequent study. Note also that the increase in the magnitude of plasma heating in the simulations with the double resolution implies that real physical processes (KHI in this case) are more important for the temperature enhancement obtained, than effects of numerical viscosity and/or resistivity. This allowed us to discuss the qualitative effect of the plasma heating, leaving quantitative estimates for forthcoming simulations.       



\section{Summary} \label{sec:summary}

In this study, we analysed transverse oscillations of a coronal loop excited by continuous monoperiodic motions of the loop footpoint. We performed a set of 3D numerical MHD simulations, considering the loop as a magnetic flux tube filled in with denser, hotter, and gravitationally stratified plasma, and analysed for the first time the response of the coronal loop to its external excitation at different frequencies.

The loop shows resonant behaviour. In particular, the loop came into resonance when the driver frequencies coincided with the frequencies of the driven standing waves established inside the loop. The amplitudes of oscillations reached the highest values at the standing wave anti-nodes, which led to the appearance of strong shear flows near the loop boundary and the development of KHI at those points. The analysis of the loop speeds showed that the amplitudes of the speed of the loop oscillating at resonant frequencies were in good agreement with theoretical expectations, although some deviation from them were identified, which was interpreted as an effect of the plasma stratification. The KHI was shown to develop for intermediate (non-resonant) frequencies as well, however, the magnitude of the flow vorticity was higher in the resonant cases.

The resonant periods for the first ($n=1$), second ($n=2$), and third ($n=3$) harmonics were detected from simulations. The ratios of the periods were slightly different to theoretical expectations for non-stratified loops, namely $1.9-2.0$ for the first-to-second harmonic ratio, and $2.9$ for the first-to-third one, because of the relatively high plasma temperature inside the loop and therefore weak stratification effect in the simulations performed. In addition, we obtained good agreement of our values for the resonant frequencies with the analytical estimates based on averaging the plasma parameters inside the loop with a weight function.

We analysed the frequency and height dependence of plasma heating by transverse waves. The positions of the maximum heating were in total agreement with those for the strength of the KHI and corresponded to the standing wave anti-nodes in the resonant cases. Comparison to the results of previous studies allowed us to conclude that the initial temperature configuration and plasma mixing played a significant role in the coronal loop heating by transverse footpoint motions. In particular, the plasma mixing due to development of KHI induced by transverse oscillations in a hotter loop resulted in the enhancement of the mean plasma temperature in the domain considered.

We should note again that the stratified loop behaviour obtained in the  simulations is in good agreement with the theoretical expectations for the gravitationally stratified case \citep{Andries2005, Verth-etal2007, ErdelyiVerth2007}. However, our study is the first in which this has been shown conclusively with three-dimensional, nonlinear numerical simulations. The findings on KHI development and height dependence of plasma heating in this study are of importance for forthcoming simulations including a broad-band driver at the loop footpoint. We plan to perform these simulations in a subsequent work.

\acknowledgments
The work was supported by the European Research Council (ERC) under the European Union’s Horizon 2020 research and innovation programme (grant agreement No.~724326). A.N.A. acknowledges support from the Russian Science Foundation under grant 17--72--10076 (simulations, analysis of results). The computational resources and services used in this work were provided by the VSC (Flemish Supercomputer Center) funded by the Research Foundation -- Flanders (FWO) and the Flemish Government -- department EWI. The results were inspired by discussions at the ISSI-Bern meeting ``Quasi-periodic Pulsations in Stellar Flares: a Tool for Studying the Solar-Stellar Connection'' and at the ISSI-Beijing meeting ``Pulsations in solar flares: matching observations and models''. 

%

\vspace{5mm}


\software{MPI-AMRVAC \citep{xia2018}
          }



\bibliographystyle{aasjournal}
\bibliography{KHI_paper_biblio}

\end{document}